\documentclass[a4paper, 12pt]{article}
\usepackage[utf8]{inputenc}
\usepackage[T1]{fontenc}
\usepackage{babel}
\usepackage{float}
\usepackage[width=170mm,height=257mm]{geometry}
\usepackage{amsmath, amsfonts, amssymb, bm}
\usepackage[pdftex]{graphicx}
\usepackage{lmodern, indentfirst, cite, xcolor}
\usepackage[squaren]{SIunits}
\usepackage[position=top,labelfont={color=blue,bf,sf}]{subfig}
\usepackage[textfont={footnotesize},labelfont={color=blue,bf,sf},labelsep=endash]{caption}
\usepackage[colorlinks=true,linkcolor=blue,citecolor=blue,urlcolor=blue]{hyperref}

\usepackage[affil-sl]{authblk}
\title{\textbf{\Large Dirac fermions collimation in heterostructures based on tilted Dirac cone materials}}
\author[a]{E.B. Choubabi\thanks{\href{mailto:choubabi.e@ucd.ac.ma}{choubabi.e@ucd.ac.ma}}}
\author[a]{B. Lemaalem}
\author[a]{M. Raggui}
\author[a]{A. Belouad}
\author[a]{R. Houça}
\author[a,b]{A. Kamal}
\author[a]{M. Monkade}
\affil[a]{LPMC Laboratory, Theoretical Physics Group, Faculty of Sciences, Chouaïb Doukkali University, 24000 El Jadida, Morocco}
\affil[b]{ISPS2I Laboratory, National Higher School of Arts and Crafts (ENSAM), Hassan II University of Casablanca, 20670 Casablanca, Morocco}
\providecommand{\pacs}[1]{\noindent \textbf{PACS numbers:} #1\\}
\providecommand{\keywords}[1]{\noindent \textbf{Keywords:} #1}
\begin{document}
%========================================================
\begin{titlepage}
    \maketitle
    \thispagestyle{empty}
    \vspace{1cm}
	\begin{abstract}
This paper aims to theoretically analyze the behavior of Dirac fermions in tilted Dirac cone material, particularly those that have diffused a barrier potential. Our results show that the degree of tilt in the $y$-direction can lead to different collimations of the Dirac fermion beams relative to the Fermi and confinement surfaces. To study the transmission probability, we exploited our  results numerically, taking into account the various configurations of the system and the different external and internal physical parameters by characterizing the behavior of fermionic transport in a proposed heterostructure. Our findings lay the groundwork for developing controllable electronic devices utilizing Dirac fermion collimation, governed by the tilt parameter, enabling precise manipulation and enhanced functionality.	
	\end{abstract}
	\vspace{5cm}
\pacs{81.05.ue, 72.80.Vp, 78.67.Wj, 71.18.+y
}
	\keywords{Graphene, potential barrier, transmission, tilted Dirac cone materials, Klein paradox, Dirac fermions, collimation, Fermi surfaces, refraction, conics }
\end{titlepage}
%========================================================
\section{Introduction}
%========================================================
Graphene, a two-dimensional carbon allotrope, consists of a single layer of carbon atoms arranged in a honeycomb hexagonal structure \cite{1,2,3,4,5,6}, renowned for its exceptional properties often dubbed as "miraculous" \cite{7}. Apart from its outstanding strength, with a tensile strength over 100 times greater than that of steel, graphene remains flexible and pliable, stretched and deformed without breaking, thereby expanding its potential applications \cite{8,9,10}.

In addition to its remarkable robustness, graphene stands out for its exceptional electrical conductivity \cite{11,12}. Electrons can flow through graphene with minimal resistance, making it an ideal material for advanced electronic devices such as transistors and conducting films. Its high thermal conductivity is also notable, facilitating efficient heat transfer, which is crucial in various thermal management applications \cite{13,14,15}.

Due to its transparency and impermeability to gases and liquids, graphene also finds applications in transparent conductive coatings, optoelectronic devices, and filtration \cite{14,17,18}. This versatility makes it relevant in fields ranging from electronics to biotechnology to energy \cite{19,20,21}.

On the other hand, tilted Dirac cone materials, such as transition metal dichalcogenide  monolayers and topological materials, exhibit unique electronic characteristics due to their deviated band structure \cite{22,23,24,25}. These materials can host tilted Dirac fermions and anisotropic transport properties, opening up new prospects in ultrafast electronic devices and spintronics\cite{28,29,30}.

The combination of graphene with these tilted Dirac cone materials promises the development of new structures, known as heterostructures, which could offer an even wider range of theoretical and practical applications \cite{31,32,33}. By combining the advantages of graphene in terms of conductivity and transparency with the unique characteristics of tilted Dirac cone materials, these heterostructures could revolutionize the fields of electronics and functional materials, thus paving the way for new opportunities in innovation \cite{34,35,36,37,38}.

Dirac fermions within heterostructures possess an inherent spin freedom, rendering them highly appealing for spin-oriented electronic applications. The collimation of these fermions offers substantial advantages across multiple technological domains \cite{380}. Firstly, in the field of spintronics, the collimation of Dirac fermions facilitates efficient manipulation and transport of spins, which is crucial for the operation of spintronic devices \cite{381}. Additionally, in the field of quantum computing, the ability to precisely control particle trajectories is essential for implementing quantum gates and processing information. Therefore, the collimation of Dirac fermions proves to be an instrumental tool in the design of quantum computing architectures \cite{382}. Moreover, the collimation of Dirac fermions holds significant potential for the development of high-speed transistors and other electronic devices offering superior performance \cite{383}. Finally, collimated Dirac fermions can also enhance the efficiency of energy conversion devices such as thermoelectric generators and photovoltaic cells. The precise regulation of electron transport provided by collimation can lead to improved conversion efficiencies and superior performance of these devices \cite{384}.

Motivated by the considerations outlined above and acknowledging the importance of Dirac fermion transport phenomena in heterostructure systems, which hold specific and significant technological implications \cite{39,40}, we initiated this study. Throughout our investigation, we meticulously scrutinize the impacts of various internal and external physical parameters of the system on transmission. These parameters include the applied electrostatic potential, the tilt factor, propagation energy, transverse and longitudinal  wave vectors, angles of incidence, and refraction. The structure of this article is as follows: Section 2 introduces the system under examination along with the theoretical model. Sections 3  presents numerical results and discussions,  while Section 4 offers concluding remarks.
%========================================================
\section{Theoretical model and formalism}\label{Sec2}
\begin{figure}[H]\centering
 	\includegraphics[scale=0.3]{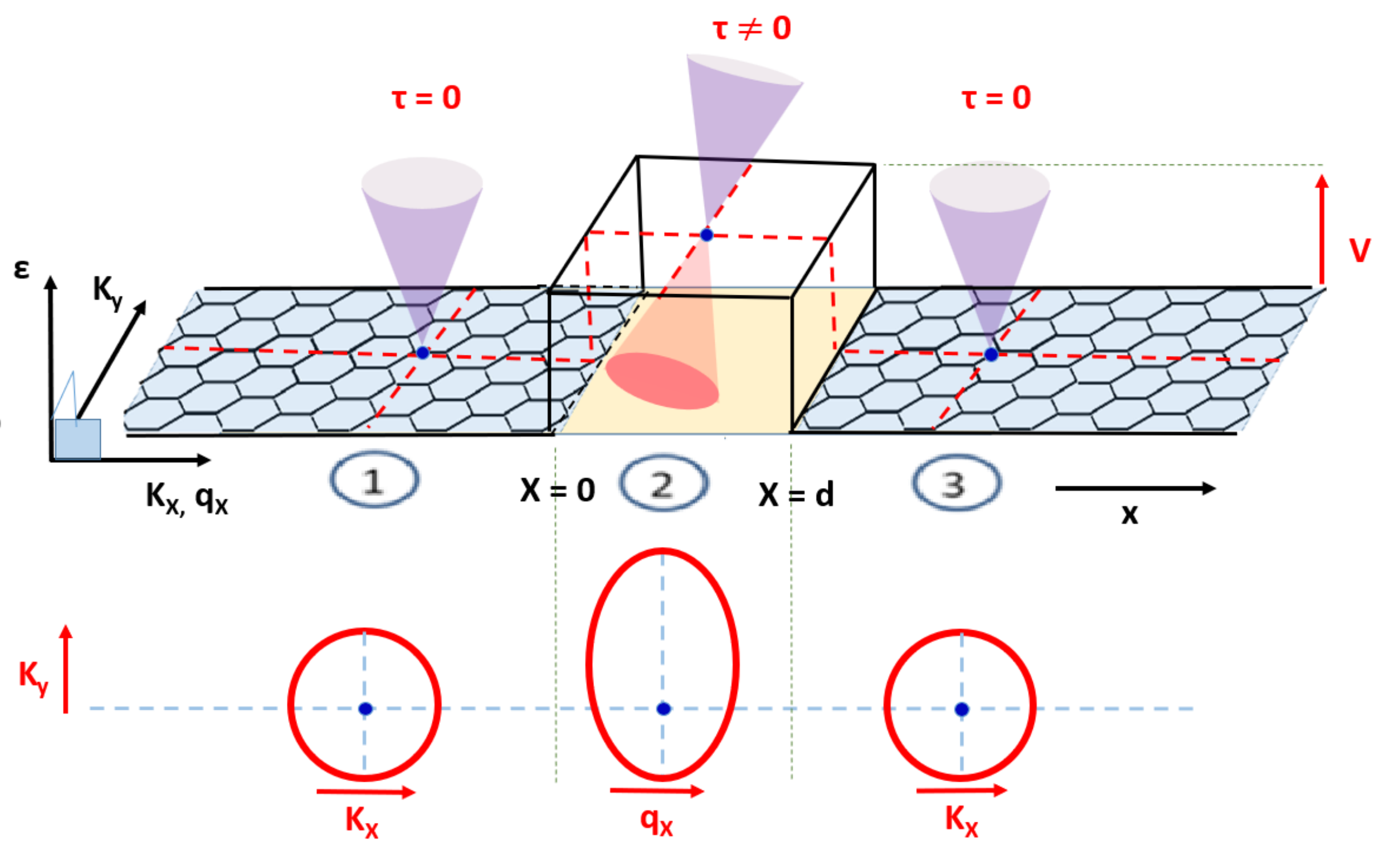}
 	\caption{(Color online) Top: An explanatory diagram illustrates a 3-region system where pristine graphene Dirac fermions encounter a potential barrier in the intermediate region containing a tilted Dirac cone material. The Dirac points are depicted in blue color.
Bottom: Fermi surfaces of each region effectively collimate the Dirac fermions during their scattering process. The foci of the Fermi surfaces are shown in blue. (The intermediate Fermi surface can be a hyperbola or a parabola depending on the value of $\tau$.)}
 	\label{Fig1}
 \end{figure}

%========================================================
In this study, we examine a heterostructure comprising three regions denoted as $\textbf{\textcircled{1}}$, $\textbf{\textcircled{2}}$, and $\textbf{\textcircled{3}}$. The input and output regions, labeled $\textbf{\textcircled{1}}$ and $\textbf{\textcircled{3}}$ respectively, consist of pristine graphene without any potential barrier. The input region extends from $-\infty$ to the junction at $x=0$, while the output region spans from the junction at $x=d$ to $+\infty$ (see Fig. \ref{Fig1}). In contrast to the adjacent regions, the intermediate region $\textbf{\textcircled{2}}$, situated between $x=0$ and $x=d$, is composed of a material characterized by a tilted Dirac cone and subject to a potential $U(x)$ defined by
\begin{equation}
U(x)=\left \{
\begin{array}{ll}
U \quad if \, \, \, 0 < x  < d \\
0 \quad otherwise \end{array}
\right.
\end{equation}

In Fig.\ref{Fig1}, a comprehensive overview of the system is presented in various regions, encompassing details about the material composition, the potential profile, dispersion relations (cones) in the reciprocal lattice, the locations of Dirac points (depicted as blue dots) that separate positive energy states contributing to electron conduction from negative energy states contributing to hole conduction. Additionally, Fermi surfaces, with the primary focus at $F(0,0)$ represented by red-colored dots, are scanned at specific Fermi energy levels for Dirac fermion scattering.

The Hamiltonian that characterizes the massless $K$-valley Dirac fermions in an isotropic tilted Dirac cone material (where $v_x=v_y=v_F$), subject to a one-dimensional potential $U(x)$ along the $x$-axis, can be explicitly expressed as follows \cite{41,42,43,44,45,46,47,48,49}:

\begin{equation}\label{eq1}
H= v_F\,\left(p_x\,\sigma_x + p_y\,\sigma_y\right)+(v_t\,p_y\, + U(x)\,)I_2
\end{equation}

In the absence of tilt, $v_F$ represents the Fermi velocity, which is set to unity, $v_F=1$. The presence of tilt introduces an additional velocity component, denoted as $v_t$. The operators $p_x=-i \hbar \partial_{x}$ and $p_y=-i \hbar \partial_{y}$ correspond to the momenta in the $x$ and $y-$directions. $\sigma_{x}$ and $\sigma_{y}$ represent the components of the Pauli matrix, and $I_2$ denotes the identity matrix. It is important to note that in this context, Planck's constant is also set to unity, denoted as $\hbar=1$. This choice simplifies the equations and allows for a dimensionless representation of physical quantities. We define the tilt parameter as $\tau=v_t/v_F=v_t$.

In the various regions, the generalized Hamiltonian can be expressed in the form of the equation \eqref{eq1}, which reduces as follows:
\begin{equation}\label{eq2}
H= \left(-i \partial_{x}\,\sigma_x + \kappa_y\,\sigma_y\right)+(\tau\,\kappa_y\, + U(x)\,)I_2
\end{equation}

In this  equation, we have considered the commutator $[H,P_{y}]=0$, which implies that the wave function in region $\textbf{\textcircled{j}}$ can be written as a tensor product $\psi_j(x,y)=\varphi_j(x)\otimes\chi(y)=\varphi_j(x)e^{i,\kappa_y,y}$, where $\kappa_y$ is the transverse wave vector. Moreover, it is important to note that the wave vector remains conserved as make between different regions. When $\tau=0$ and $U(x)=0$, we obtain the pristine graphene Hamiltonian in regions $\textbf{\textcircled{1}}$ and $\textbf{\textcircled{3}}$ \cite{48,50,51}.

By solving the eigenvalue equation for each region, we obtain the eigenspinors and eigenvalues as given by equations (\eqref{eq2}-\eqref{eq3}). To simplify the equations, we have introduced the following quantities, where these parameters are dimensionless. The parameters $ \kappa_x $ and $ \mathfrak{q}_x $ correspond to the longitudinal wave vectors of pristine graphene and the tilted Dirac cone material, respectively.
\begin{equation}
    \epsilon=\dfrac{E d}{\hbar\,v_F},\quad V=\dfrac{U d}{\hbar\,v_F},\quad k_{x}= \kappa_{x} d,\quad k_{y}= \kappa_{y} d, \quad \text{and}\quad q_{x}=\mathfrak{q}_{x} d
\end{equation}

By separating the variables, in each region the spinor is written as
\begin{equation}
\psi_j(x,y)=\left(
 \begin{array}{cc}
 \psi_j^+(x) \\ \psi_j^-(x)
 \end{array}
 \right) e^{i\,k_y\,y},\qquad j=1,2,3.
\end{equation}
In region $\textbf{\textcircled{1}}$ ($x < 0$), the eigenspinor is
\begin{equation}\label{eq2}
 \psi_1=\left(
 \begin{array}{cc}
 1 \\s_1 z_1
 \end{array}
 \right) e^{i(k_x\,x+k_y\,y)} +r\left(
 \begin{array}{cc}
 1 \\ -s_1z_1^{-1}
 \end{array}
 \right)e^{i(-k_x\,x+k_y\,y)}
\end{equation}
with
\begin{equation}
z_1= \frac{k_x + i\,k_y}{\sqrt{k_x^2 + k_y^2}},\qquad s_1=\text{sign}(\epsilon)
\end{equation}

where $r$ is the reflection amplitude, and the sign function $s_1 = \text{sign}(\epsilon)$ serve to indicate the conduction and valence bands. It is relatively easy to derive the corresponding dispersion relation.

\begin{equation}
	\varepsilon = s_1\sqrt{k_x^2 + k_y^2}.
\end{equation}

The longitudinal wave vector $k_x$ can be expressed as follows:
\begin{equation}\label{sa1}
	 k_x = \sqrt{\varepsilon^2 - k_y^2}.
\end{equation}

In the region $\textbf{\textcircled{2}}$ ($ 0 < x < d $), the eigenspinor can be expressed as
\begin{equation}
	\psi_2= a\left(
	\begin{array}{cc}
		1 \\s_2 z_2
	\end{array}
	\right) e^{i(q_x\,x+k_y\,y)} +b\left(
	\begin{array}{cc}
		1 \\ -s_2 z_2^{-1}
	\end{array}
	\right)e^ {i(-q_x\,x+k_y\,y)}
\end{equation}
where $z_2$ and $s_2$ are written as follows:
\begin{equation}
	z_2= \frac{q_x + i\,k_y}{\sqrt{q_x^2 + k_y^2}},\qquad s_2= sign(\epsilon-\tau\,k_y - V)
\end{equation}
given that $a$ and $b$  are the amplitudes corresponding to forward and backward propagation, respectively. The corresponding energy dispersion is
\begin{equation}
	\epsilon=V+\tau\,k_y+s_2\sqrt{q_x^2 + k_y^2}
\end{equation}
The longitudinal wavevector $q_x$ can be represented in the following manner:
\begin{equation}\label{sa2}
	 q_x= \sqrt{(\epsilon-\tau\,k_y -V)^{2} - k_y^2}
\end{equation}
In region $\textbf{\textcircled{3}}$ ($ x > d $), the spinor that propagates with the same wave vector $k_1$ as in region  $\textbf{\textcircled{1}}$ can be expressed as follows:

\begin{equation}\label{eq3}
	\psi_3= t \left(
	\begin{array}{cc}
		1 \\ s_1 z_1
	\end{array}
	\right) e^{i(k_x\,x+k_y\,y)}
\end{equation}
where $t$ is the transmission amplitude.

The $y-$axis component of the wave vector, denoted as $k_y$, remains conserved throughout all encountered regions during diffusion. Beyond the barrier, it can be represented as:
\begin{equation}\label{sal1}
 k_y = \varepsilon\,\sin\theta
\end{equation}
 Conversely, within the barrier, it is defined as:
\begin{equation}\label{sal2}
  k_y = |\epsilon- \tau\,k_y-V |\sin\phi
\end{equation}
 In external regions, the perpendicular component of the wave vector is represented as $k_x = \epsilon\cos\theta$, where $\theta$ denotes the incident angle, determined by $\theta=\arctan (k_y/k_x)$.

The transverse wave vector conservation  allows us to equalize the two limbs as follows :
\begin{equation}\label{000}
 \varepsilon \sin\theta=|\epsilon- \tau\ k_y-V |\sin\phi
\end{equation}
This latter equation is equivalent to Snell's Law (also known Ibn-Sahl's Law). According to this law, for a given pair of media, the ratio of the sines of the  incidence angle  and the refraction angle  equals the ratio of the two media refractive indices $n_j$ \cite{52,53,54,55}, which enables us to express it as:

 \begin{equation}\label{001}
 \frac{\sin\theta}{\sin\phi}=\frac{|\epsilon- \tau\ k_y-V |}{ \varepsilon}=\frac{n_{2}}{n_1}
\end{equation}
This law enables us to predict a comparison between the two regions in terms of refringence.

To deepen our understanding of collimation phenomena, it is crucial to thoroughly examine the configuration of the Fermi surface in each region $\textbf{\textcircled{j}}$. Therefore, in the extreme regions, we can elaborate on the Cartesian equation that characterizes the isoenergetic Fermi surfaces as follows:

 \begin{equation}\label{eq4}
	k_x^2 + k_y^2  =\epsilon^2
\end{equation}
The specific form of this equation will depend on the particular parameters of the system under consideration.

In the case of the middle region, the Fermi surface can be characterized by the following Cartesian equation \cite{42}:
\begin{equation} \label{eq5}
	q_x^2 + \left(1-\tau^2\right)\,k_y^2 +2\,\tau\,k_y\left(\epsilon-V\right) - \left(\epsilon-V\right)^2 =0
\end{equation}

It is important to highlight that the equation \eqref{eq4} represents a specific scenario derived from equation \eqref{eq5} when substituting $\tau=0$ and $V=0$. In the literature, the characteristics of tilted Dirac cone materials undergo a systematic classification into four distinguishable phases, determined by the value of $\tau$: the untilted phase ($\tau=0$), type I phase ($0<\tau<1$), type II phase ($\tau>1$), and type III phase ($\tau=1$) \cite{44,47}.

\begin{figure}[H]\centering
\hspace{-10mm}
\subfloat[Untilted]{
        \includegraphics[scale=0.10]{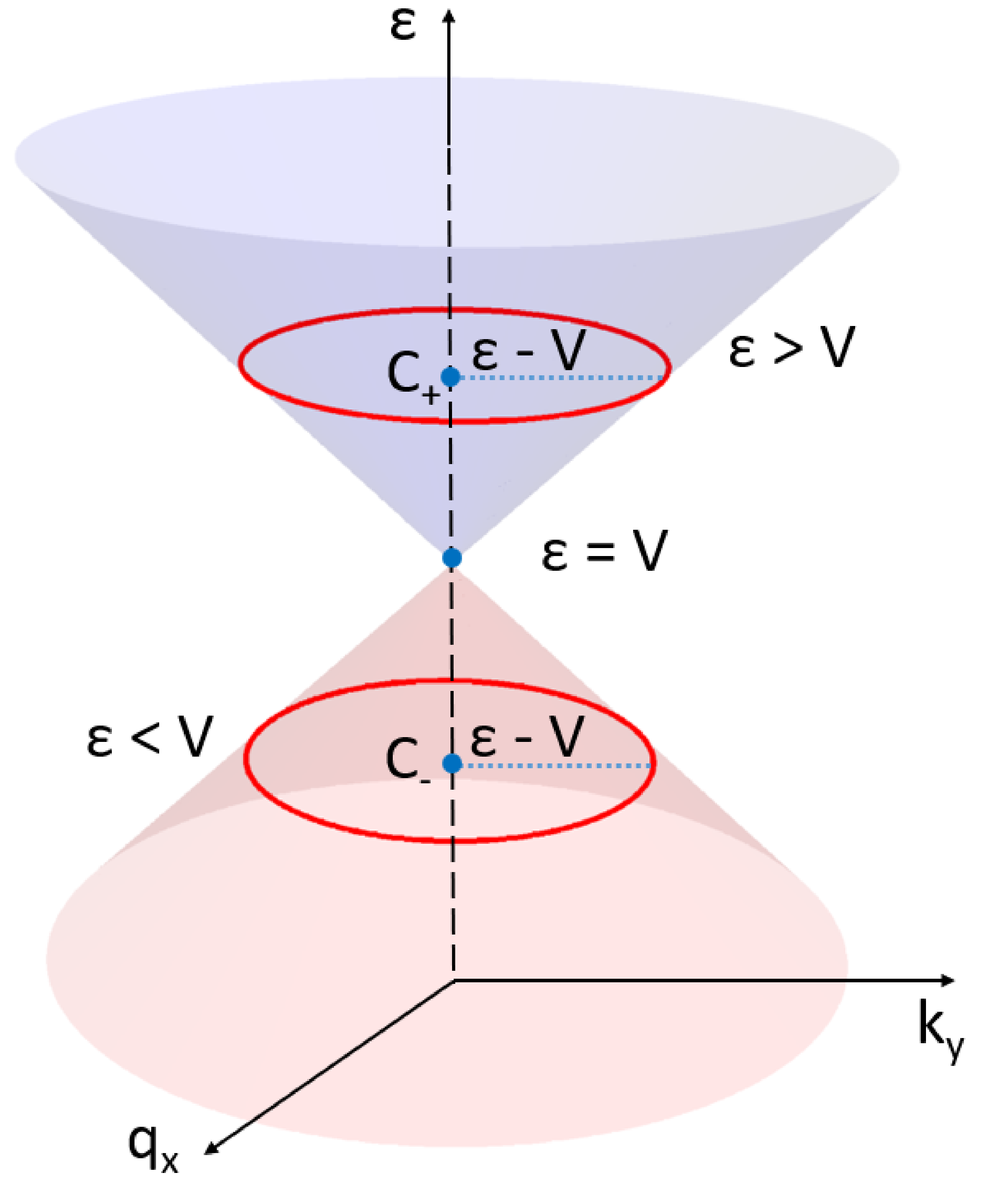}
        \label{fig02:SubFigA}
    }\hspace{-5mm}
	\subfloat[Type $I$]{
        \includegraphics[scale=0.10]{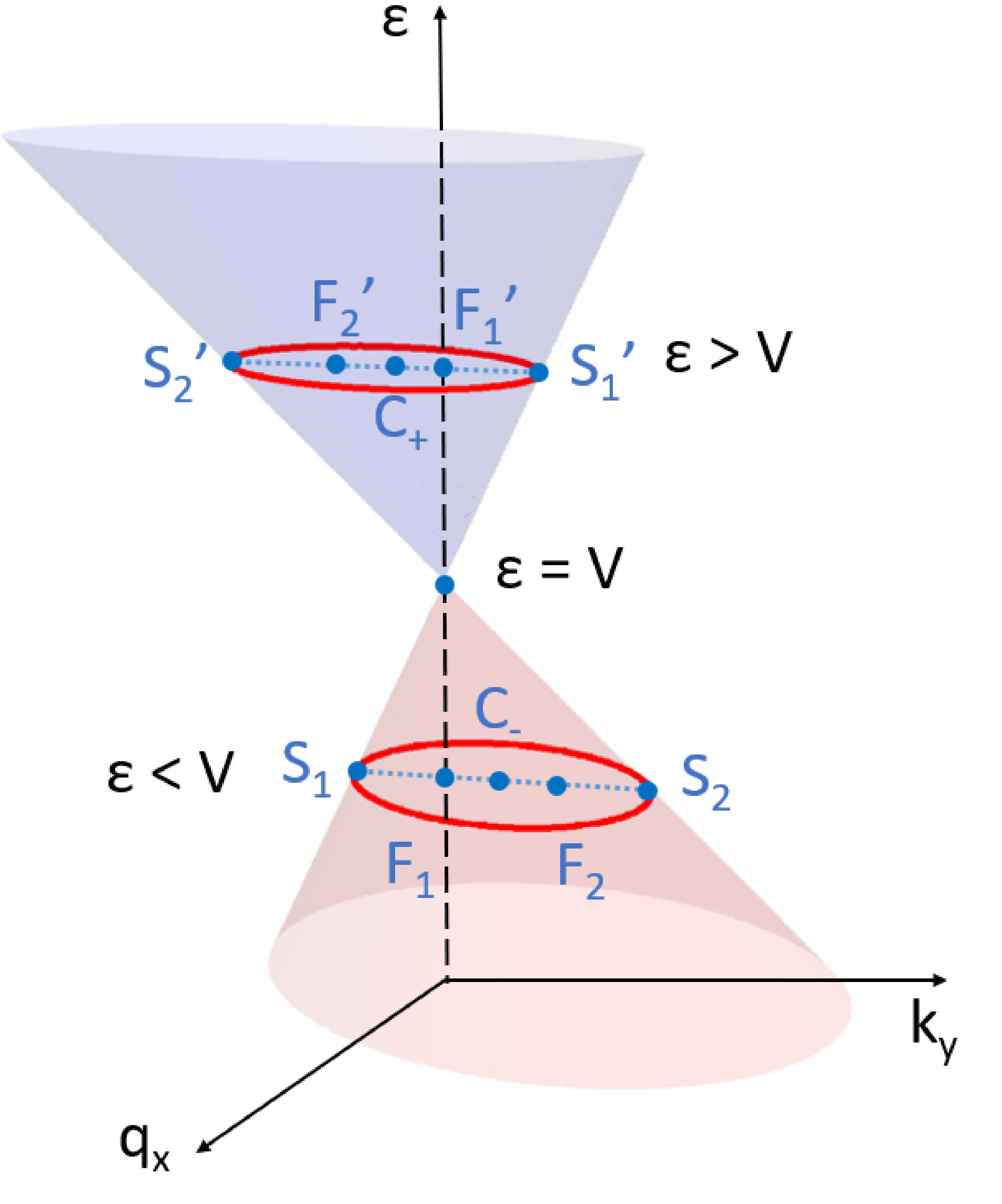}
        \label{fig02:SubFigB}
    }\hspace{-5mm}
	\subfloat[Type $II$]{
        \includegraphics[scale=0.11]{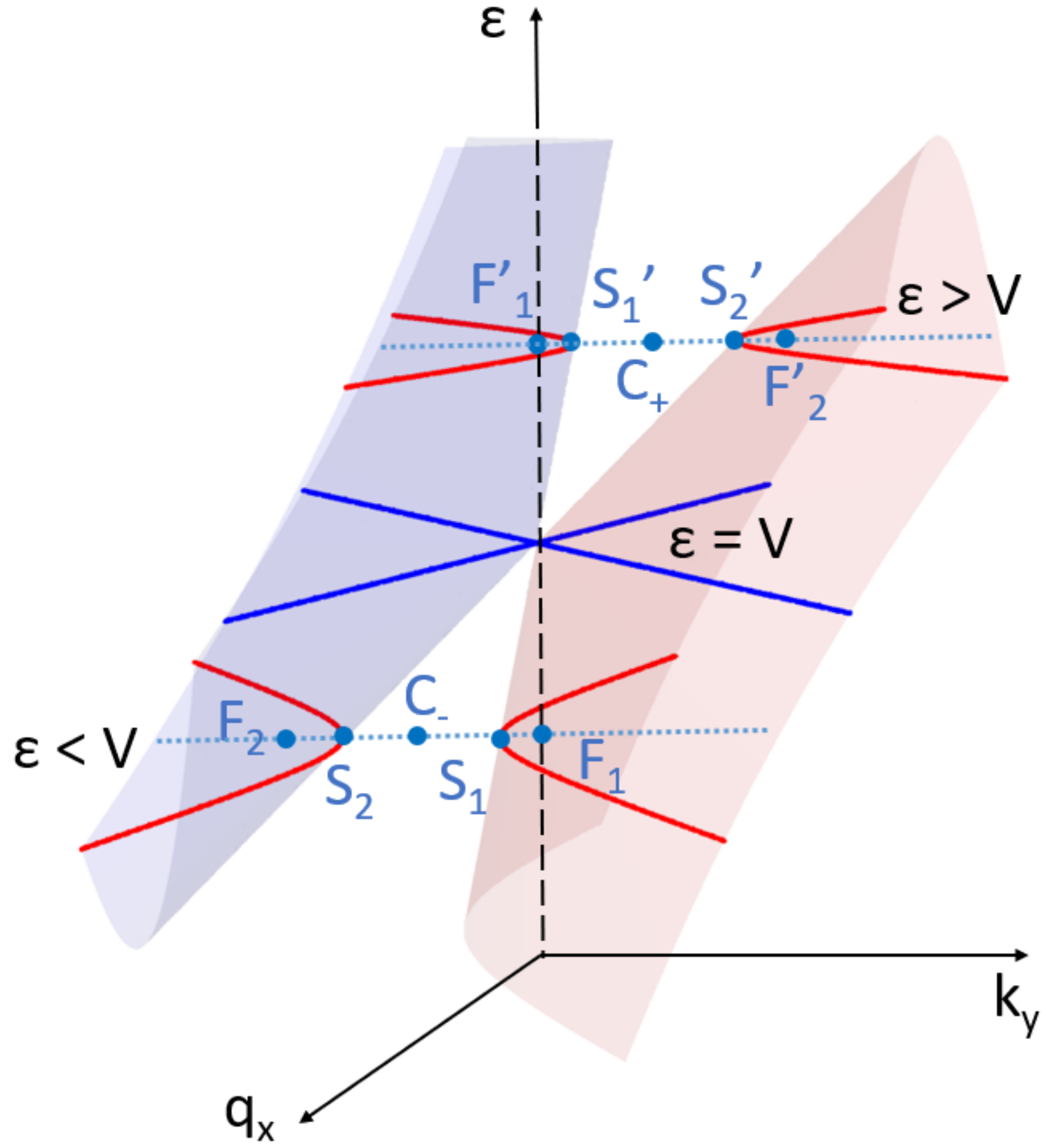}
        \label{fig02:SubFigC}
    }\hspace{-3mm}
	\subfloat[Type $III$ ]{
        \includegraphics[scale=0.10]{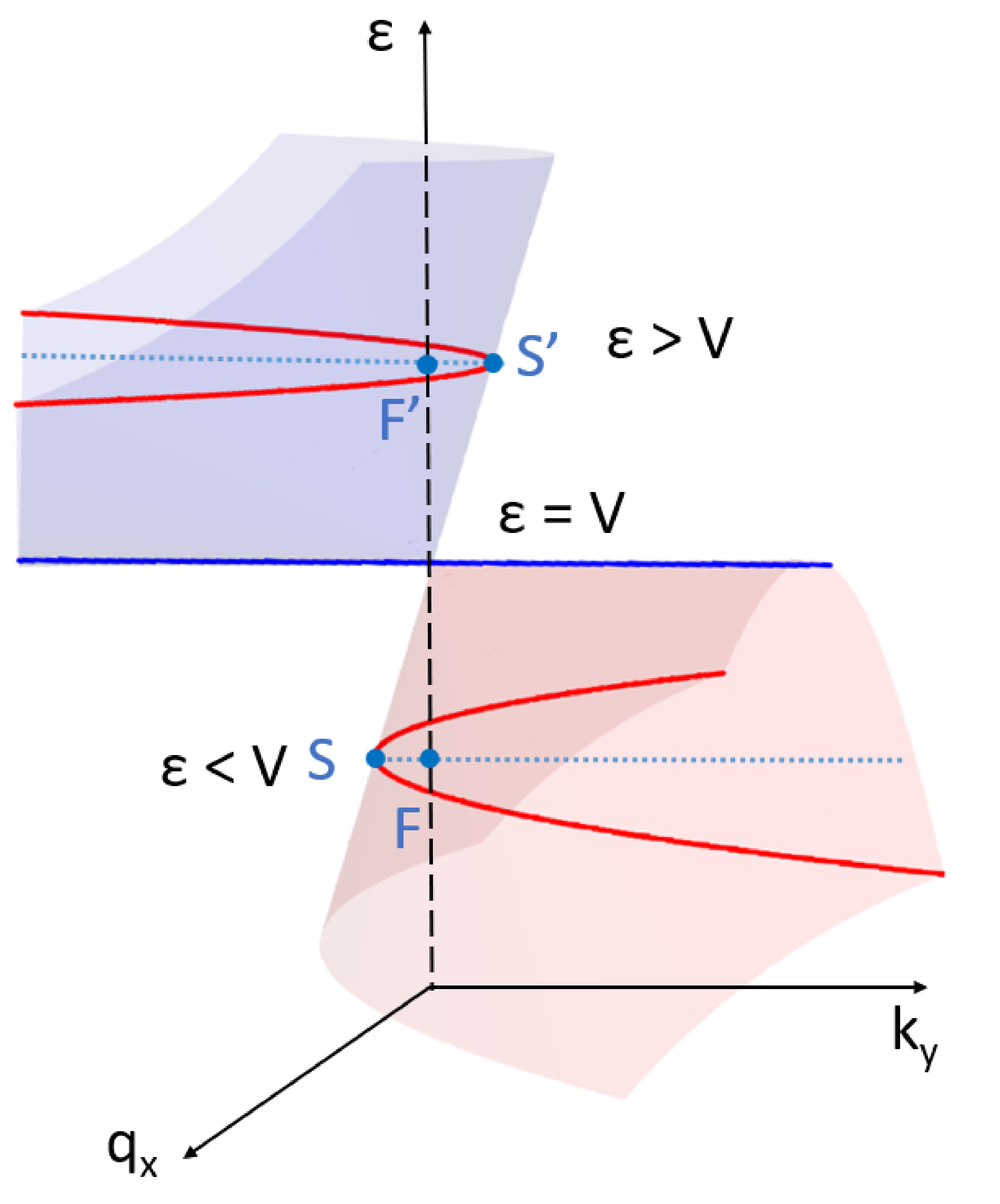}
        \label{fig02:SubFigD}
    }
	\caption{(Color online) Various types of tilted Dirac cones, in reciprocal space, exhibiting  Fermi surfaces for different energy values. the localization indication  of geometric parameters for the conical curves obtained at each energy, as well as the Dirac points generated at the intersection of the conduction and valence bands.}
	\label{fig02}
\end{figure}

Before delving into the exploration of various types of inclined cones, let's elucidate Fig. \ref{fig02}, which offers a comprehensive view of data related to these cone variations. It provides a detailed graphical representation of Fermi surfaces in reciprocal space, capturing different predicted cones, each characterized by a unique configuration. The tilt physical parameter, ($\tau$), plays a decisive role in determining the nature of each cone.

Upon closer inspection of the figure, it becomes apparent how variations in  by a unique configuration. The crucial physical parameter of deviation, $\tau$ influence the geometry and properties of the associated Fermi surfaces. This visual representation serves as a crucial foundation for comprehending the relationship between the tilt parameter and the shape of cones, laying the groundwork for an in-depth study of material properties associated with these structures in reciprocal space.
Within the figure, one can observe the cone locations, Fermi surface energy dependence, and the localization of geometric parameters of the obtained conics. It also highlights the tilt effect on the Fermi surface at the energy level $\epsilon = V$, showcasing various scenarios such as Dirac points, lines formed by sets of Dirac points, and intersecting lines formed by Dirac points \cite{44,55}. These scenarios will be discussed in more detail later.

In pristine graphene, the energy dispersion relation, commonly referred to as the band structure, exhibits a linear behavior in the vicinity of the Dirac point (Fig. \ref{fig02:SubFigA}) with the condition $V=0$). This linearity is characterized by the absence of an energy gap, indicating that charge carriers within the material behave as massless Dirac fermions \cite{550}. The linear relationship near the Dirac point is a distinctive feature of graphene's electronic structure, playing a crucial role in determining the unique electronic and transport properties associated with this remarkable material \cite{551}. The absence of a gap in the energy dispersion allows for exceptional electronic mobility and unconventional transport phenomena, contributing to the distinct behavior of charge carriers in pristine graphene.

In the first phase (untilted) (Fig. \ref{fig02:SubFigA}), where $\tau = 0$, the Fermi surface is confined within a circular boundary. This circular boundary follows the same equation as the Fermi surface of pristine graphene under the influence of a potential $V$. If the potential $V$ is set to zero, the region is governed by equation \eqref{eq4}. In regions $\textbf{\textcircled{1}}$ and $\textbf{\textcircled{3}}$, the circle has its center at $C(0,0)$ and a radius of $\epsilon$.

In the second phase (type I) (Fig. \ref{fig02:SubFigB}), characterized by \(0 < \tau < 1\), the Fermi surface adopts the shape of an ellipse, characterized by the following equation \cite{42,56,57}:
\begin{equation}
    \left(\frac{k_y -\dfrac{ \tau \left(V-\epsilon\right) }{ 1-\tau^2} }{\dfrac{\epsilon-V}{1-\tau^2} }\right)^2+\left(\frac{q_x}{\dfrac{\epsilon-V}{\sqrt{1-\tau^2}}}\right)^2 =1
\end{equation}

Now, let's delve into the geometric properties of this ellipse:
The center of the ellipse is situated at \(C_{\pm}\left(0, \dfrac{\pm\tau(\epsilon-V)}{1-\tau^2}\right)\).
The ellipse has a major axis along the \(k_{y}\)-axis with a length of \(\dfrac{\mid\epsilon-V\mid}{1-\tau^2}\).
The minor axis extends along the \(q_x\)-axis and has a length of \(\dfrac{\mid\epsilon-V\mid}{\sqrt{1-\tau^2}}\).

Where conduction takes place within the valence band, and the energy condition \((\epsilon < V)\) suggests the mobilization of holes, the two foci are located at \(F_1(0, 0)\) and \(F_2\left(0, \dfrac{2\tau|\epsilon -V|}{1-\tau^{2}}\right)\), and the two vertices are positioned at \(S_1\left(0, -\dfrac{|\epsilon -V|}{1+\tau}\right)\) and \(S_2\left(0, \dfrac{|\epsilon -V|}{1-\tau}\right)\). Contrary to conduction by electrons with \((\epsilon > V)\), the two foci are located at \(F_1'(0, 0)\) and \(F_2'\left(0, \dfrac{-2\tau|\epsilon -V|}{1-\tau^{2}}\right)\), and the two vertices are positioned at \(S_1'\left(0, \dfrac{|\epsilon -V|}{1+\tau}\right)\) and \(S_2'\left(0, -\dfrac{|\epsilon -V|}{1-\tau}\right)\).

In the third case (type II) (Fig. \ref{fig02:SubFigC}), characterized by $\tau > 1$, the Fermi surface takes the form of a pair of hyperbolas, as described by the equation \cite{42,56,57}:

\begin{equation}
\left(\frac{k_y - \dfrac{\tau(\epsilon-V)}{\tau^2-1}}{\dfrac{\epsilon-V}{\tau^2-1}}\right)^2 - \left(\frac{q_x}{\dfrac{\epsilon-V}{\sqrt{\tau^2-1}}}\right)^2 = 1
\end{equation}

These hyperbolas are centered at $C_{\pm}\left(0, \dfrac{\pm\tau|\epsilon-V|}{\tau^2- 1}\right)$.
The various geometric parameters of the hyperbolas delimiting the isoenergetic Fermi surfaces for both the conduction and valence bands are determined as follows: $C_{\pm}\left(0, \dfrac{\pm\tau|\epsilon-V|}{\tau^2- 1}\right)$, $F_1(0, 0)=F_1'(0, 0)$, \(S_1\left(0, \dfrac{-|\epsilon -V|}{1+\tau}\right)\), \(S_1'\left(0, \dfrac{|\epsilon -V|}{1+\tau}\right)\), \(S_2\left(0, \dfrac{-|\epsilon -V|}{\tau -1}\right)\), \(S_2'\left(0, \dfrac{|\epsilon -V|}{\tau -1}\right)\), \(F_2\left(0, \dfrac{-2\tau|\epsilon -V|}{\tau^{2} -1}\right)\) and \(F_2'\left(0, \dfrac{2\tau|\epsilon -V|}{\tau^{2} -1}\right)\)

In the  final case (type III) (Fig. \ref{fig02:SubFigD}), $\tau=1$,  the Fermi surface takes the form of a parabola that satisfies the equation \cite{42,56,57}:

\begin{equation}
	q_x^2 =2 (V-\epsilon) \left( k_{y}-\dfrac{(\epsilon-V)}{2}\right)
\end{equation}

In this case, the two parabolas delimiting the isoenergetic Fermi surfaces have geometric parameters expressed as follows: $F(0, 0)=F'(0, 0)$, \(S\left(0, \dfrac{-|\epsilon -V|}{2}\right)\), and \(S'\left(0, \dfrac{|\epsilon -V|}{2}\right)\).

For Fermi surfaces exhibiting various configurations of tilted Dirac cones, depending on the propagation energy, there is an observed symmetry between the surfaces of positive energy $(\epsilon > V)$ and those of negative energy $(\epsilon < V)$. This central symmetry occurs at $(0,0,V)$, positioned within the plane that separates the conduction and valence bands. As a result, the positive and negative surfaces intersect within this plane, leading to the generation of Dirac points.

The tilted effect on Fermi surfaces at the energy $\epsilon = V$ results in distinct Dirac point configurations:
In Fig. \ref{fig02:SubFigA}, for this illustrated configuration, we have a single Dirac point located at the symmetry center of the two conduction and valence cones at the corners presented by the point $k$ in the first Brillouin zone. Here, the Fermi surface exhibits a punctual structure coinciding with the Dirac point $(0,0,V)$, indicating linear dispersion in the vicinity of this point.
Also in Fig. \ref{fig02:SubFigB}, we have a Dirac point at $(0,0,V)$ similarly to Fig. \ref{fig02:SubFigA}. It shares the same characteristics with a point-like structure and linear dispersion near the Dirac point.
Fig. \ref{fig02:SubFigC} shows two crossed lines formed by a set of Dirac points: In this configuration, the Fermi surface presents two intersecting lines composed of several Dirac points. Each line represents the locus of Dirac points at energy $\epsilon = V$, forming a crossing pattern within the Brillouin zone.
Fig. \ref{fig02:SubFigD} indicates a line formed by a set of Dirac points: Here, the Fermi surface forms a line consisting of a series of Dirac points at the energy $\epsilon = V$. This line-like structure indicates the linear dispersion of electronic states along a specific direction in the Brillouin zone, with each Dirac point marking a significant transition in the electronic band structure.

In summary, the tilted effect on the Fermi surface at the energy $\epsilon = V$ yields various configurations of Dirac points, each representing unique features of the electronic band dispersion in the material .

\begin{figure}[H]\centering
\hspace{-10mm}
	\subfloat[$\tau=0$]{\includegraphics[scale=0.19]{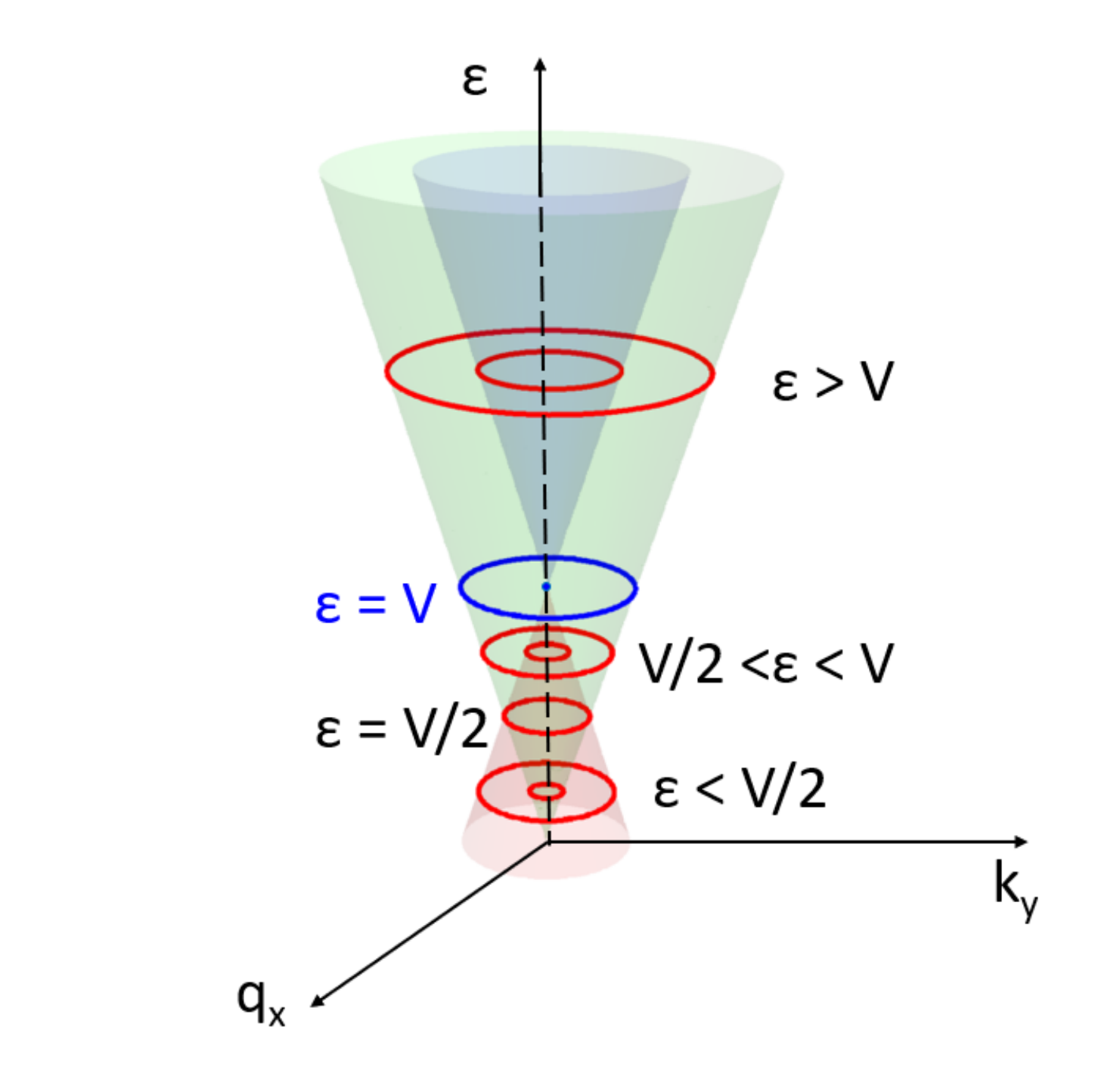}\label
{fig03:SubFigA}}
	\hspace{-3mm}
	\subfloat[$0<\tau<1$ ]{\includegraphics[scale=0.17]{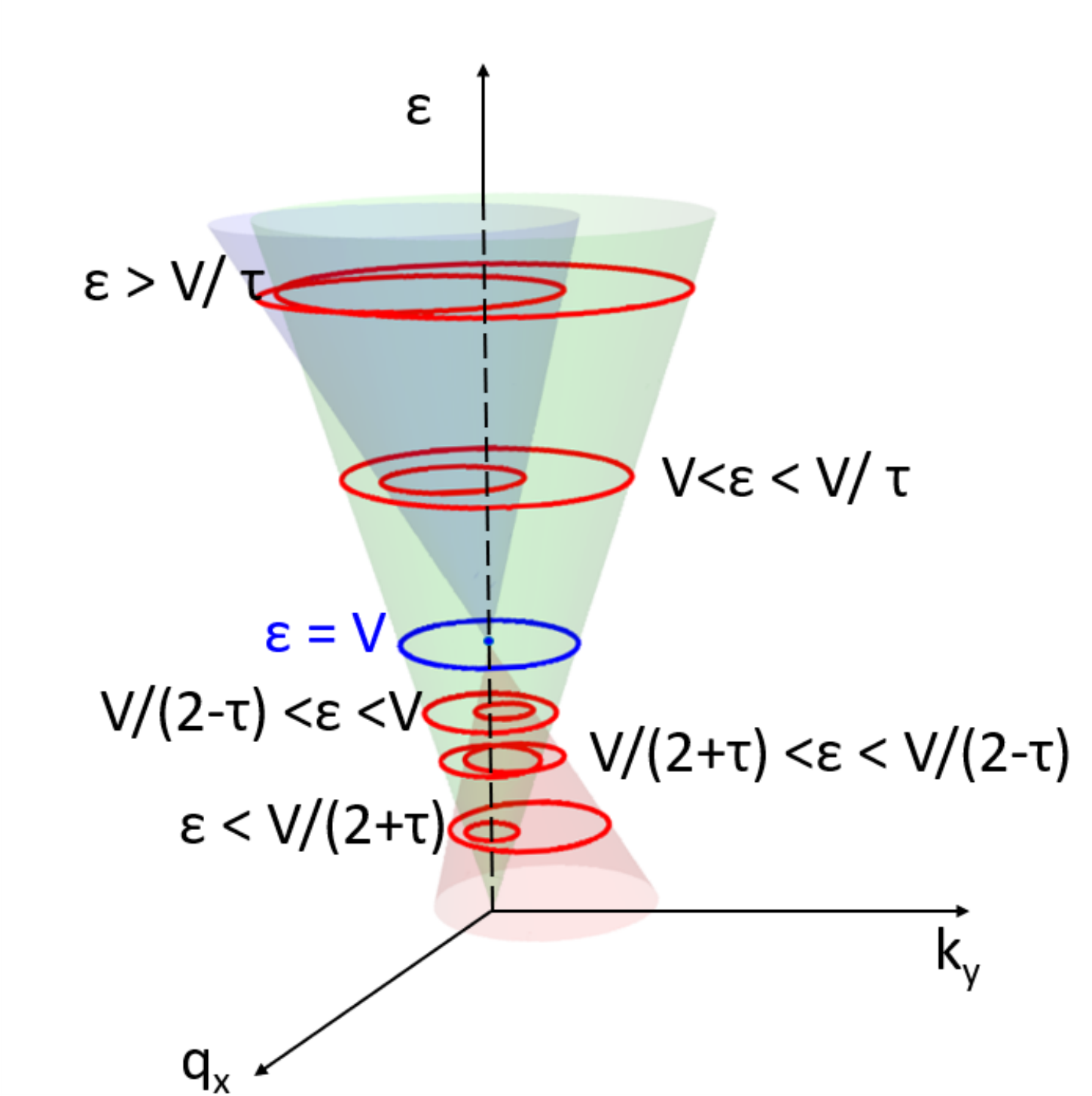}\label
{fig03:SubFigB}}
	\hspace{-2mm}
	\subfloat[$\tau = 1$]{\includegraphics[scale=0.11]{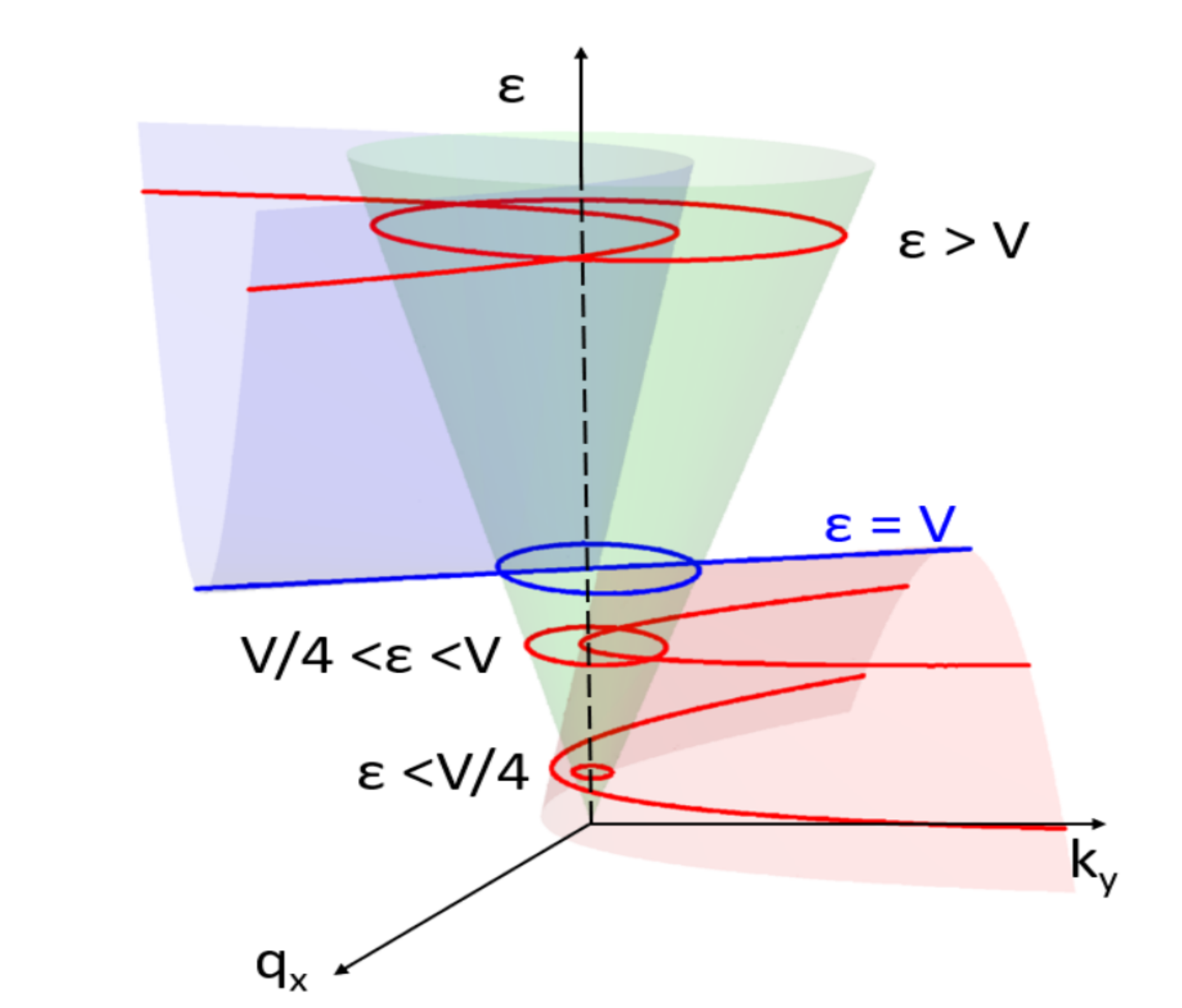}\label
{fig03:SubFigC}}
\hspace{-4mm}
	\subfloat[$\tau > 1$]{\includegraphics[scale=0.095]{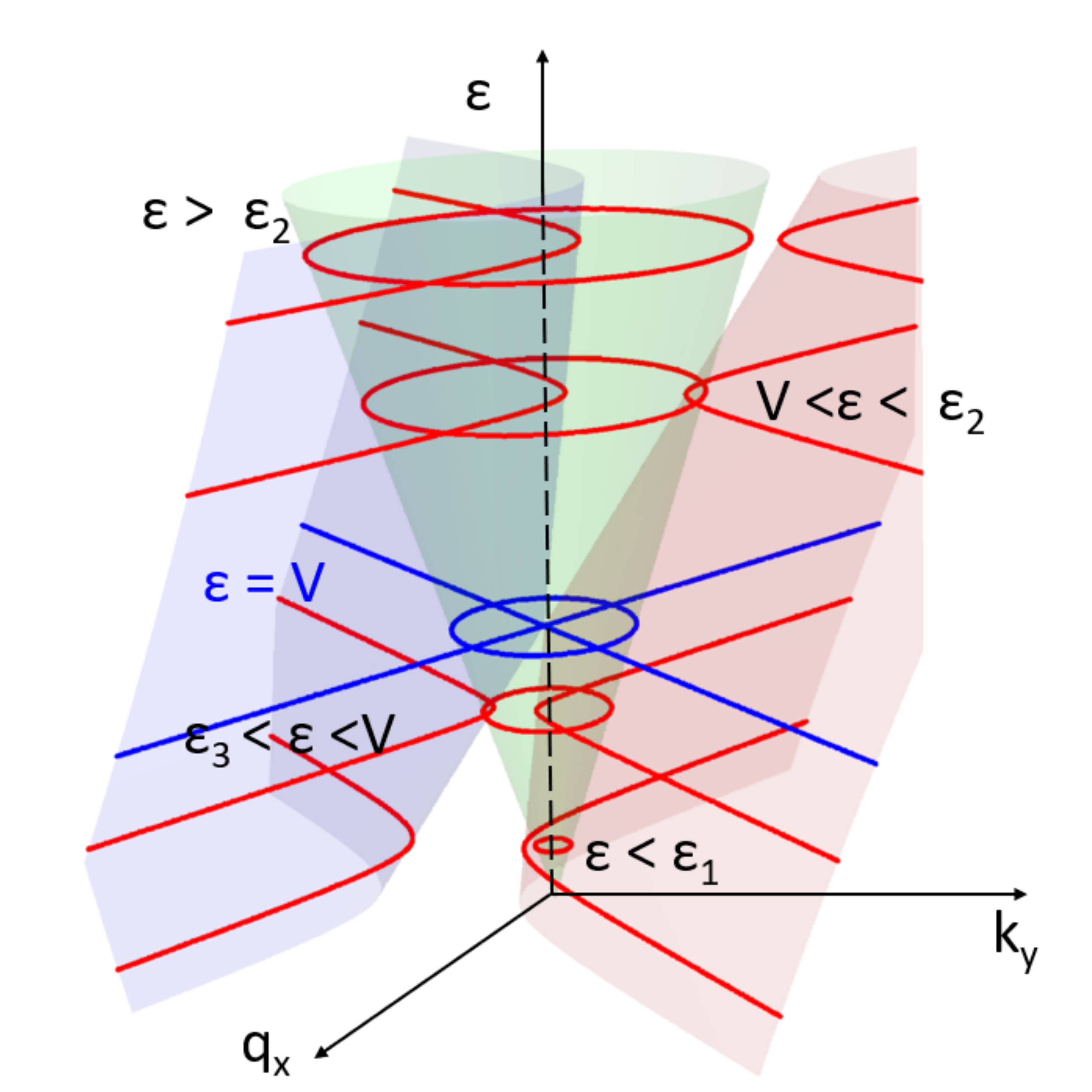}\label
{fig03:SubFigD}}
	\caption{(Color online) Evolution of the active surfaces, due to collimation during the diffusion of Dirac fermions, as a function of the propagation energy
Evolution of the active surfaces, due to collimation during the diffusion of Dirac fermions, as a function of the propagation energy $\epsilon$ for different values of $\tau $. }
	\label{fig03}
\end{figure}

Fig. \ref{fig03} offers a detailed exposition on the evolution of the active surface within the system \cite{42,45,47}. It portrays the active surface as an amalgamation of Fermi surfaces originating from diverse regions denoted as $\textbf{\textcircled{j}}$. This overlap is a consequence of collimation, where predicted surfaces align, facilitating their fusion into the active surface. Significantly, the active surface delineates
the permitted transmission region during diffusion at $\epsilon$ propagation energy, thus guiding transmission dynamics. This figure provides nuanced insights into the emergence and characteristics of active surfaces, elucidating their pivotal role in transmission during diffusion at $\epsilon$ propagation energy.

For Dirac fermion beams to propagate through the system's regions, they must conform to respective Hamiltonians, necessitating simultaneous scanning of isoenergetic surfaces across different regions. This scanning generates active surfaces, which essentially represent overlaps of isoenergetic surfaces traversed during propagation. Understanding these active surfaces is paramount for comprehending the propagation dynamics of Dirac fermion beams across the system's varied zones.

The Fig. \ref{fig03} shows the interpenetrations of cones that delineate the permitted and forbidden transmission zones \cite{50,58,59}, depending on the tilted parameter. Specific domains are observed based on the energy and transverse wave vector.
In the figure, the green cone (conduction band) represents the entry and exit regions of the system, while the two blue and red cones represent the intermediate region. The blue cone (conduction band) corresponds to the positive energy levels of the electrons, while the red cone (valence band) corresponds to the negative energy levels of the holes. The energy levels are calculated and given by: $\epsilon=V/(2+\tau)$, $\epsilon=V/(2-\tau)$ and $\epsilon=V/\tau$.

In Fig.  \ref{fig03:SubFigA}, the interpretation of the cones provides two specific energy levels, $\epsilon=V/2$ and $\epsilon=V$, which determine three distinct domains of propagation energies. For domain $A_{1}=]0,V/2[$, the intersection of the Fermi surfaces results in an active surface in the form of a circle that just touches the contour of the green cone. Thus, this active surface is defined by the surface of the green cone, which is entirely contained within the Fermi surface of the red cone. Conduction within this active surface occurs through electrons in both extreme regions and through holes in the intermediate region.
In contrast, for domain $B_{1}=]V/2,V[$, the active surface is that of the red cone, which collimates the incident beam of Dirac fermions and is completely included in the surface of the green cone. The union of domains $A_{1}\cup B_{1}$ generates a lozenge in the $(\epsilon, k_{y})$-plane, which is characterized by the tunneling effect of Dirac fermions. In this case, the heterostructure behaves like an $NPN$ junction \cite{60,61}.
For domain $C_{1}=]V,\infty[$, the active surface coincides with the blue cone, where we will only have electron conduction in all regions of the system.
For $\epsilon=V/2$, the two Fermi surfaces are equal and give the same active surface with a radius of $\epsilon=V/2$, and for $\epsilon=V$,  the green cone Fermi surface  forms a circle with a radius of $\epsilon=V$, but those of the blue and red cones are reduced to a Dirac point.

When $\tau$ takes a value strictly between $0$ and $1$, the cones in the intermediate region undergo a tilt without touching the plane corresponding to $\epsilon=V$ (Fig. \ref{fig03:SubFigB}). This configuration of cone interpenetration generates, this time, four ordered energy levels: $\epsilon_{1}=V/(2+\tau)$, $\epsilon_{2}=V/(2-\tau)$, $\epsilon=V$, and $\epsilon_{3}=V/\tau$.
These four ordered energy levels generate five defined domains. In the first domain $A_{2}=]0,\epsilon_{1}[$, the active surface is circular and similar to that of domain $A_{2}$ in the previous case. However, this surface is entirely contained within the ellipse generated by the tilted red cone, with its left focus coinciding with the center of the predicted circle. Moreover, both foci focus at $k_{y}=0$ corresponding to normal incidence at the system's entrance.
If $\epsilon=\epsilon_{1}$, the active surface is a circle originating from the green cone and belonging to the ellipse obtained from the red cone. The left vertex of the ellipse coincides with the edge of the circle, and the left focus coincides with the center of the circle. In the second domain $B_{2}=]\epsilon_{1},\epsilon_{2}[$, the active surface is a common section between the circle and the ellipse. The center of the circle remains coincident with the left focus of the ellipse, while the left vertex $S_{1}$ lies within the circle and the right vertex $S_{2}$ extends beyond the circle. When $\epsilon=\epsilon_{2}$, vertex $S_{2}$ is on the circle.
For energies within the domain $C_{2}=]\epsilon_{2},V[$, the active surface is an ellipse entirely contained within the circle. For $\epsilon=V$, the situation is entirely similar to that of the untilted cone. In domain $D_2=]V,\epsilon_{3}[$, the active surface is similar to that of domain $C_2$, except that there is an enlargement and a change in the focus of the ellipse. This time, the right focus coincides with the center of the circle. For $\epsilon = \epsilon_{3}$, the ellipse enlarges along with the circle, and the left vertex $S'_2$ lies on the circle, while $S_2$  is inside the circle. Beyond $\epsilon > \epsilon_{3}$, there is an overflow of from the circle. The union $A_{2}\cup B_{2}\cup C_{2}$  forms a quadrilateral which represents a deformed rhombus, where the tunneling effect of Dirac fermions occurs, and the system behaves similarly to an $NPN$ junction.

In the figure depicting the tilted Type III cone ($\tau=1$) in the intermediate region (Fig. \ref{fig03:SubFigC}), the configuration represents an extreme case of the previous configuration when $\tau \rightarrow 1$. The blue and red cones tilted by $\pi/2$ to reach just the plane $\epsilon=V$. This modification will result in $\epsilon_{3}=V$  and the elliptical Fermi surfaces being replaced by other parabolic ones. The energy domains will be reduced to three: $A_{3}=]0,V/4[$, $B_{3}=]V/4,V[$ and $C_{3}=]V,\infty[$ . The Dirac point will be replaced by a set of Dirac points forming a line along $\epsilon=V$, and the quadrilateral will have its upper edge coinciding with the generated Dirac points.

In the Fig. \ref{fig03:SubFigD}, where $\tau > 1$ (type II), the ordered energy levels are: $\epsilon_{1} = V/(2+\tau)$, $\epsilon_{3} = V/\tau$, $\epsilon = V$, and $\epsilon_{2} = V/(2-\tau)$. In the intermediate region, the Fermi surfaces undergo a significant shape change due to the the cones tilt  greater than $\pi/2$. This  tilt causes the Fermi surfaces to take a hyperbolic shape, characterized by two distinct branches symmetrical to each other. These branches are disjoint, meaning they do not cross or overlap.
Overall, the hyperbolic shape of the Fermi surfaces in the intermediate region reflects the complex interplay between the  cone tilt and the resulting changes in the electronic structure of the material. For energies $\epsilon \in A_{4}=]0,\epsilon_{1}[$, the circle representing the Fermi surface of the in and out regions of the system is entirely situated inside the right branch of the hyperbolic surface. At $\epsilon=\epsilon_{1}$, the circle of the active surface touches the vertex $S_{1}$ of the right branch, with its focus $F_{1}$ coinciding with the circle. In domain $B_{4}=]\epsilon_{1},\epsilon_{3}[$, the active surface is the intersection between the circle of the green cone and the right branch of the red cone. The part of the circle that extends beyond the branch is forbidden. Conduction in this active surface occurs through holes in the intermediate region. For $\epsilon=\epsilon_{3}$, the circle touches the vertex $S_2$ of the corresponding the blue cone left branch. Concerning $\epsilon \in C_{4}=]\epsilon_{1},V[$, the circle generates two active surfaces, one with the right branch and another with the left branch of the blue cone. In the left branch, conduction through the intermediate zone is via electrons. This means that for the same energy level, two types of conduction of electrons and holes can appear respectively on the blue and red cones, and this depends on the angle of incidence. At $\epsilon =V$, the Fermi surfaces of the red and blue cones reduce to a set of Dirac points in two intersecting lines form, while that of the green cone reduces to a circle with radius $\epsilon =V$. The active surfaces in domain $D_{4}=]V,\epsilon_{2}[$ are similar to those in domain $D_{4}$, except that the roles of the two branches are swapped and there is an enlargement of these surfaces. Similarly, the active surfaces in $E_{4}=]\epsilon_{2},\infty[$ are also similar to those in $B_{4}$. The lozenge extends beyond plane $\epsilon=V$ and gives rise to a quadrilateral bounded by point $\epsilon_{2}/(2-\tau)$, and when $\tau=2$ is reached, the upper edge extends to infinity.

A detailed theoretical study of specific energy levels created by the overlapping of cones in different configurations allowed us to identify the corresponding points of these levels in the $(\epsilon, k_y)$-plane. The entirety of these determined points is: $ (k_1,\epsilon_1 )=(-V/(2+\tau),V/(2+\tau) )$, $ (k_2,\epsilon_2 )=(V/(2-\tau),V/(2-\tau) )$ and $ (k_3,\epsilon_3 )=(-V/\tau,V/\tau )$. We observe that the coordinates of these points are influenced by both the height of the barrier and the tilt parameter. This dependency results in a variation of these points for each configuration $(V,\tau)$. In other words, changes in the barrier height or the tilt parameter lead to corresponding shifts in the positions of these points in the $(\epsilon, k_y)$-plane. Therefore, the specific characteristics of the barrier and tilt parameter play a crucial role in determining the energy levels associated with the overlapping of cones.

To calculate the transmission coefficient accurately, we employ the transfer matrix method, a powerful tool widely used in quantum mechanics. This method relies on maintaining the continuity of the wavefunction  across different regions within the system. At the interfaces, located at \(x = 0\) and \(x = d\), the spinor wavefunctions must remain continuous. Mathematically, this continuity condition is expressed in terms of transfer matrices, denoted as $\mathcal{M}_i(x)$ $(i=1,2)$, corresponding to the transformations between different regions.

At the interface \(x = 0\), the transfer matrix \(\mathcal{M}_1\) describes the transition from one region to another. This matrix encapsulates the behavior of the wavefunction as it enters the heterostructure. Similarly, at \(x = d\), the transfer matrix \(\mathcal{M}_2\) characterizes the transformation as the wavefunction exits the heterostructure, allowing us to understand how the wavefunction evolves throughout the system \cite{50,59}.

By solving the equations governed by these transfer matrices, we can determine the transmission coefficient, which represents the probability of an incident particle passing through the system. Through this detailed analysis, we gain insights into the quantum mechanical behavior of particles as they propagate through the heterostructure, providing valuable information for various applications in nanotechnology and quantum devices.

\begin{equation}
	\mathcal{M}_i(x)=\begin{pmatrix}
 e^{i\,k_i\,x} &  e^{-i\,k_i\,x}\\
 z_i\,e^{i\,k_i\,x} & -s_i\,z_i^{-1}\,e^{i\,k_i\,x}
	                 \end{pmatrix}
\end{equation}
is transfer matrix that connect the $i$-th region wavefunction to the ($i + 1$)-th region wavefunction. Then, over the tilted Dirac cone material, the full transfer matrix can be expressed as
\begin{equation}
	\left(
	\begin{array}{cc}
		1 \\ r
	\end{array}
	\right)=M\left(
	\begin{array}{cc}
		t \\ 0
	\end{array}
	\right) =M_1^{-1}(0). M_2(0). M_2^{-1}(d). M_1(d)
	\left(
	\begin{array}{cc}
		t \\ 0
	\end{array}
	\right)
\end{equation}
and then
\begin{equation}
	M=\begin{pmatrix}M_{11} & M_{12}\\ M_{21} & M_{22}\end{pmatrix}
\end{equation}
As a result, the transmission coefficient is given by
\begin{equation}
	t=\frac{1}{M_{11}}
\end{equation}
At this stage, using the incident and transmitted current densities $J_i$ and $J_t$ to calculate the transmission probability $T=\tfrac{J_t}{J_i}$. Indeed, we get the associated current density
\begin{equation}
	J=e\, v_F\, \psi^+ \sigma_x\, \psi
\end{equation}
corresponding to our system, leading to the following transmission probability
\begin{equation}
	T=|t|^2.
\end{equation}

\section{Results and discussion}\label{Sec3}
%========================================================

In this section, we will conduct a comprehensive examination of our proposed system, providing detailed insights into its unique features and thoroughly exploring how the transmission probability varies with the system's different physical parameters. To accomplish this, we will utilize the analytical results obtained previously, which will be further clarified through numerical analysis.
Our primary focus in this study will be on investigating the impact of the tilted cone parameter, denoted by $\tau$, along the $y$ direction within reciprocal space. Specifically, we will analyze its effects on the behavior of Dirac fermions within pristine graphene when encountering a tilted Dirac cone material serving as a barrier. We will delve into the specifics of how this parameter influences the transmission properties and other relevant characteristics of the heterostructure under examination.
Furthermore, we will meticulously examine the transmission and refraction of the regions as functions of various internal and external parameters of the system. These parameters include the transverse moment $k_y$, the longitudinal moment $k_x$, and the incidence angle $\theta$ of  Dirac fermions beam at the junction separating the first region from the second. Additionally, we will explore the Fermi surfaces and active surfaces resulting from the collimation phenomenon as functions of incident energy.
It's important to note that if $\tau < 0$, all cones will be tilted in the opposite direction with a negative angle. Due to this symmetry, we will solely concentrate on the scenario where $\tau > 0$.

\begin{figure}[H]\centering

\subfloat[$\tau=0$]{\includegraphics[scale=0.095]{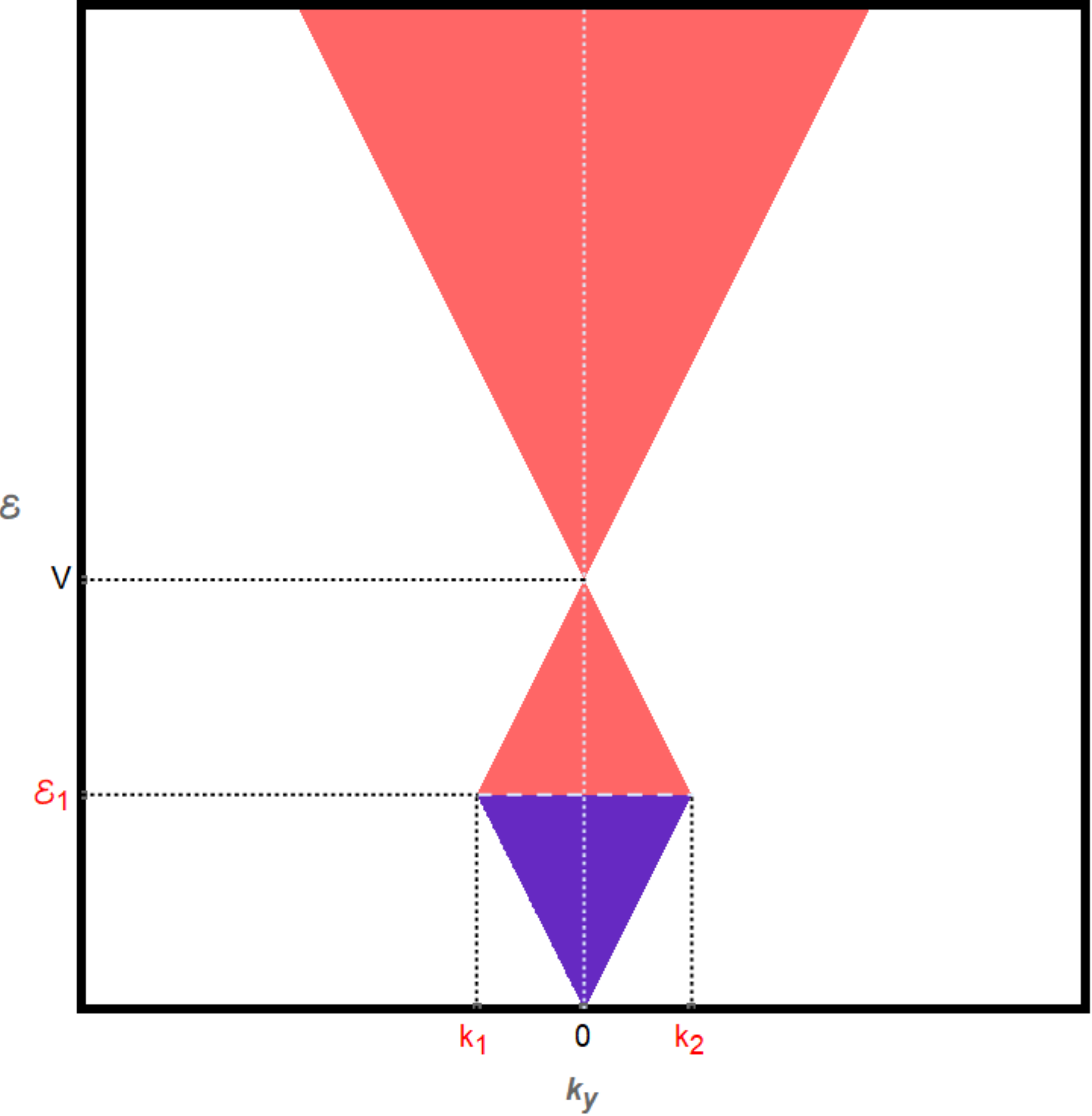}\label{TEnThetaPlusPi1:SubFigA}}
	\hspace{-0.1mm}
	\subfloat[ $0<\tau<1 $]{\includegraphics[scale=0.095]{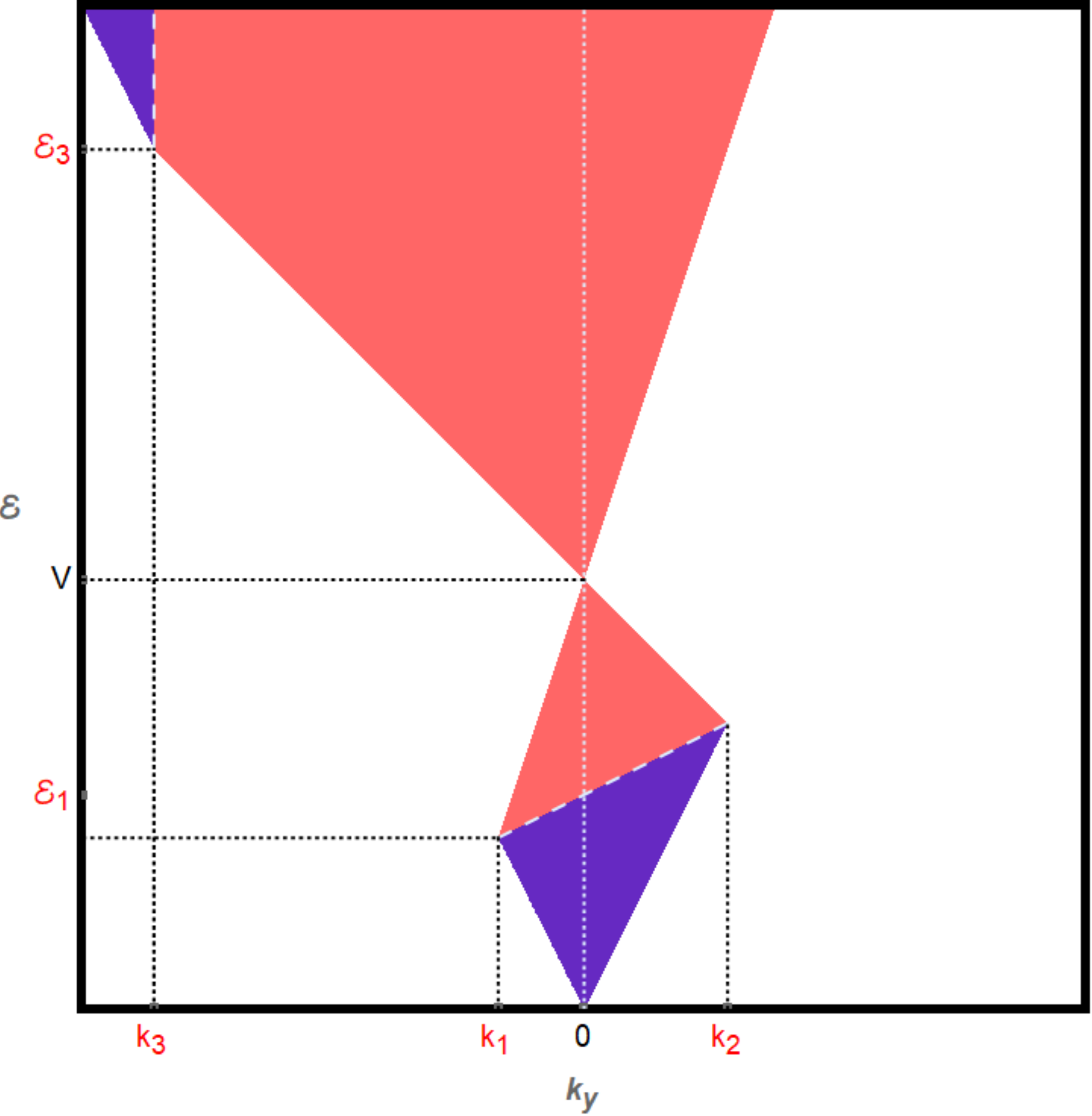}\label{TEnThetaPlusPi1:SubFigB}}
	\hspace{-1mm}
	\subfloat[$\tau=1$]{\includegraphics[scale=0.095]{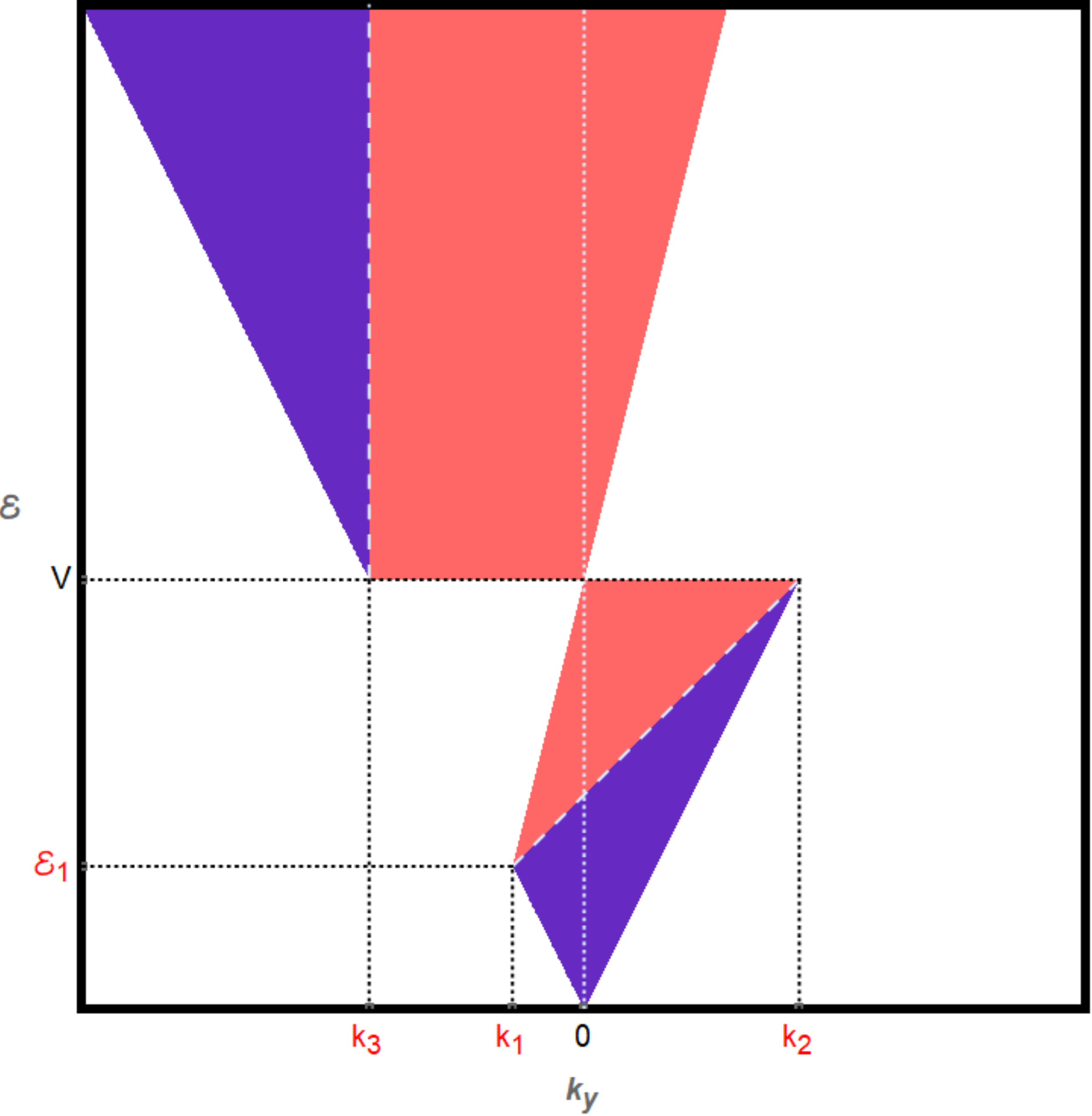}\label{TEnThetaPlusPi1:SubFigC}}
	\hspace{-0.1mm}
	\subfloat[$\tau>1$ ]{\includegraphics[scale=0.095]{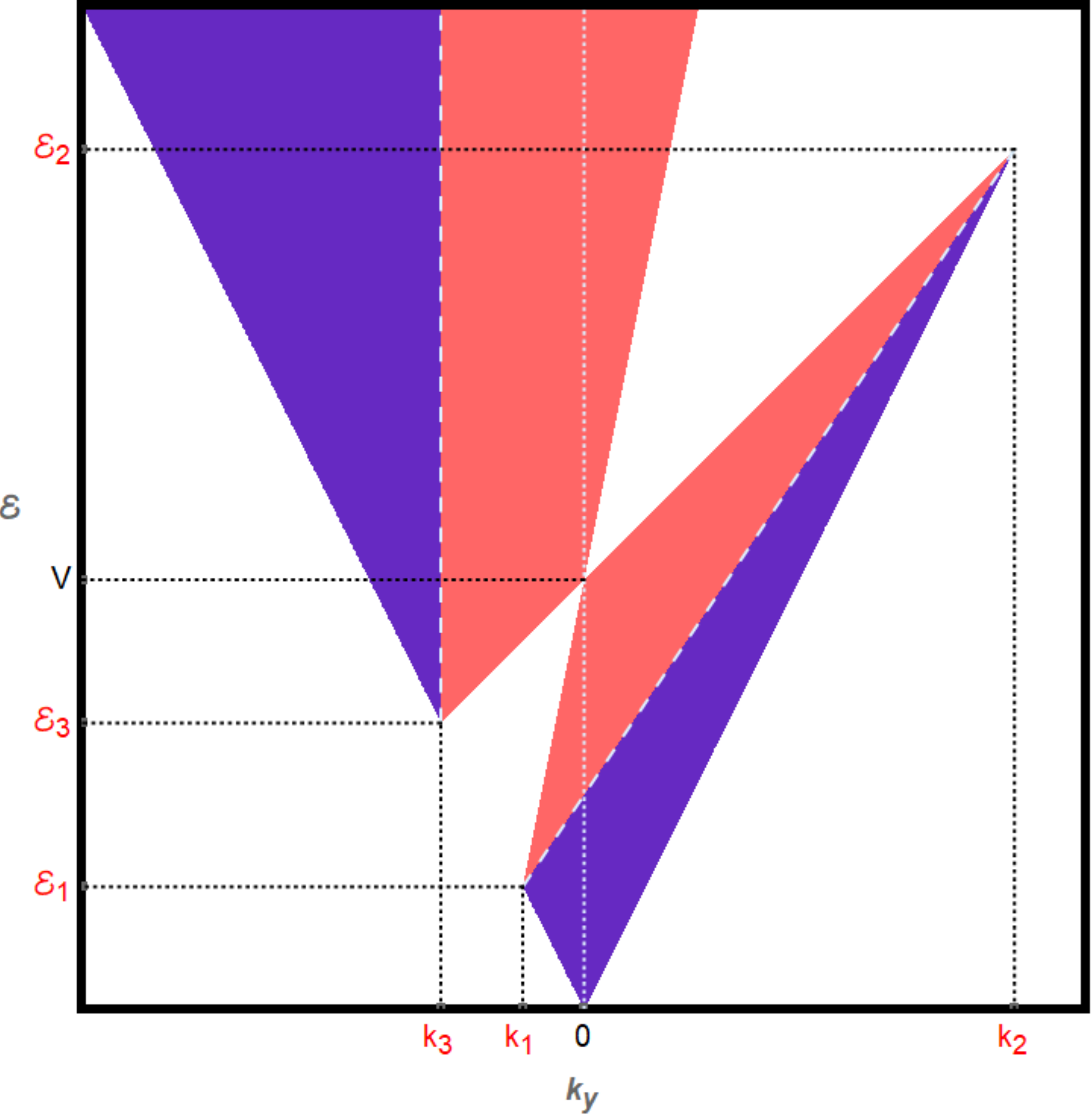}\label{TEnThetaPlusPi1:SubFigD}}
	\hspace{-3mm} \subfloat[$\tau=0$]{\includegraphics[scale=0.097]{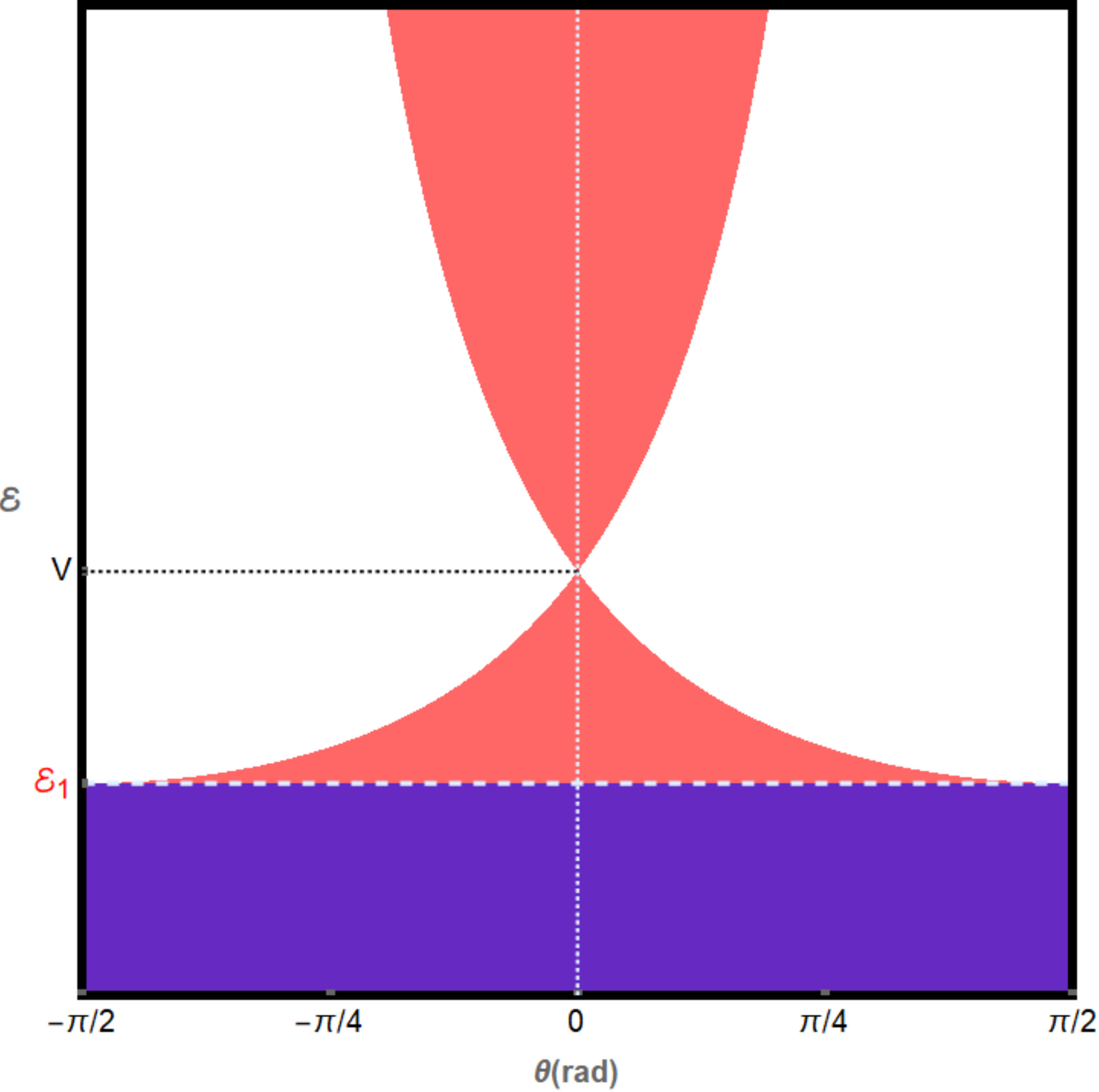}\label{TEnThetaPlusPi1:SubFigE}}
	\hspace{-0.1mm}
	\subfloat[ $0<\tau<1 $]{\includegraphics[scale=0.095]{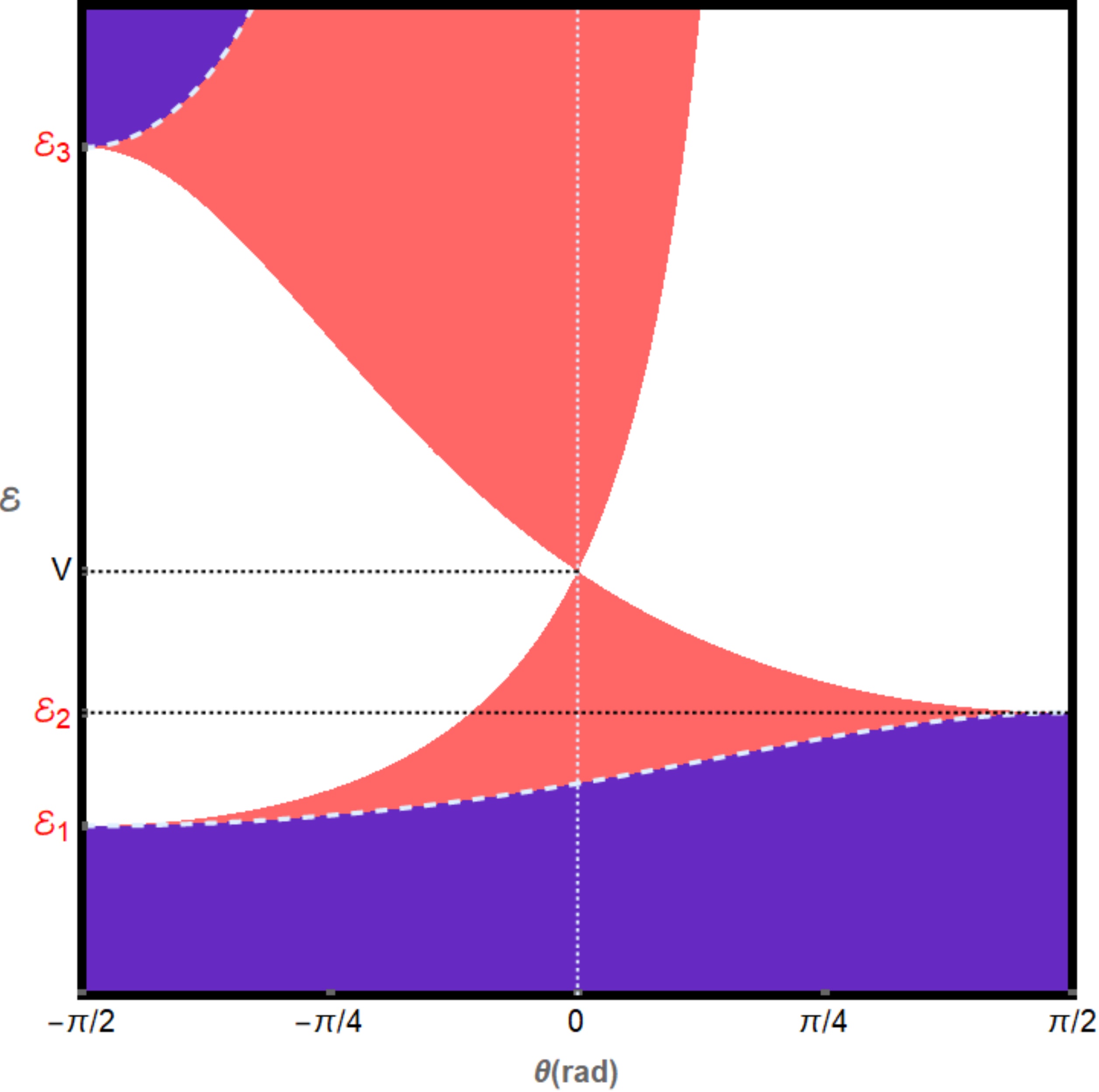}\label{TEnThetaPlusPi1:SubFigF}}
	\hspace{-1mm}
	\subfloat[$\tau=1$]{\includegraphics[scale=0.095]{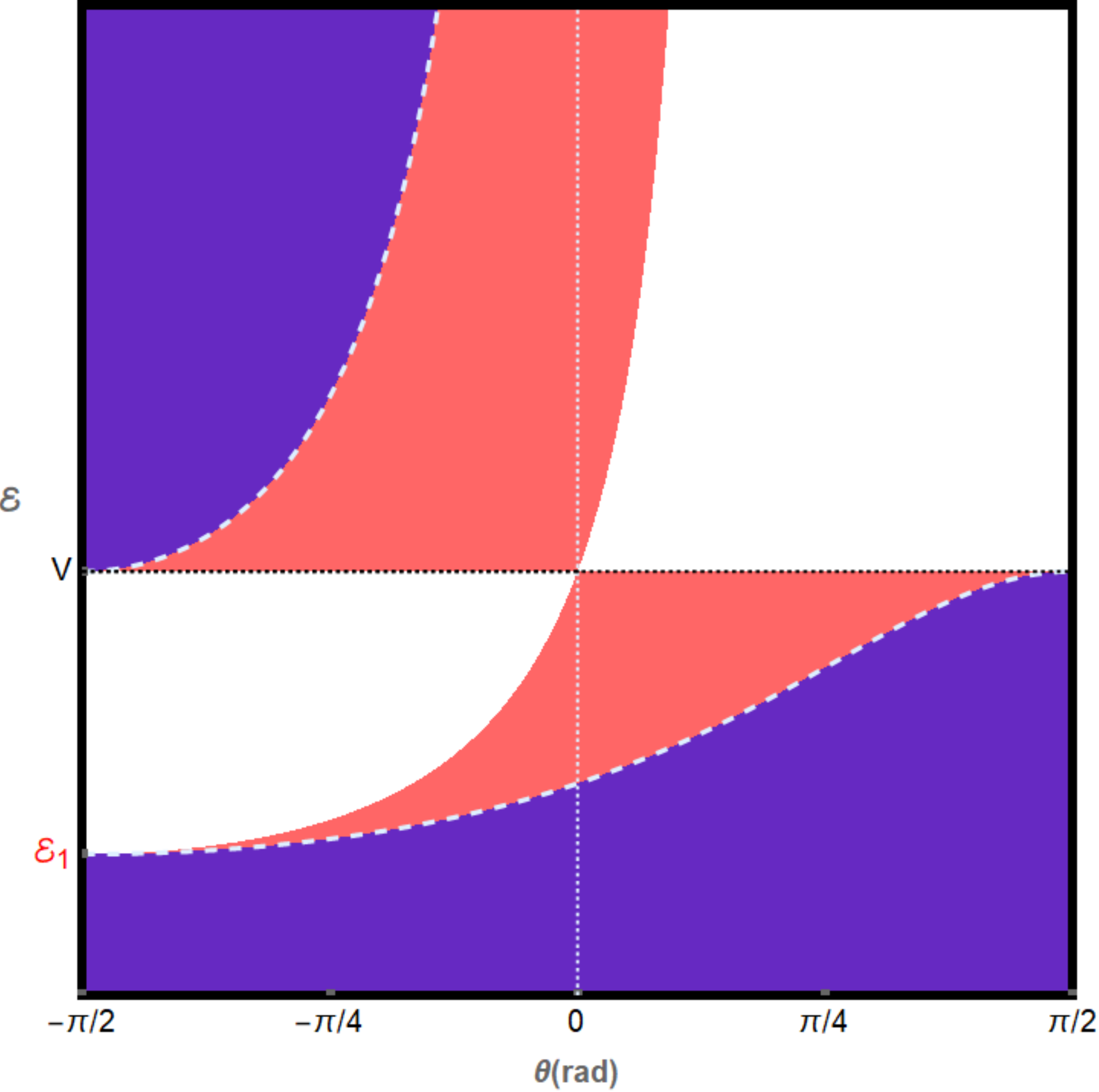}\label{TEnThetaPlusPi1:SubFigG}}
	\hspace{-0.1mm}
	\subfloat[$\tau>1$ ]{\includegraphics[scale=0.095]{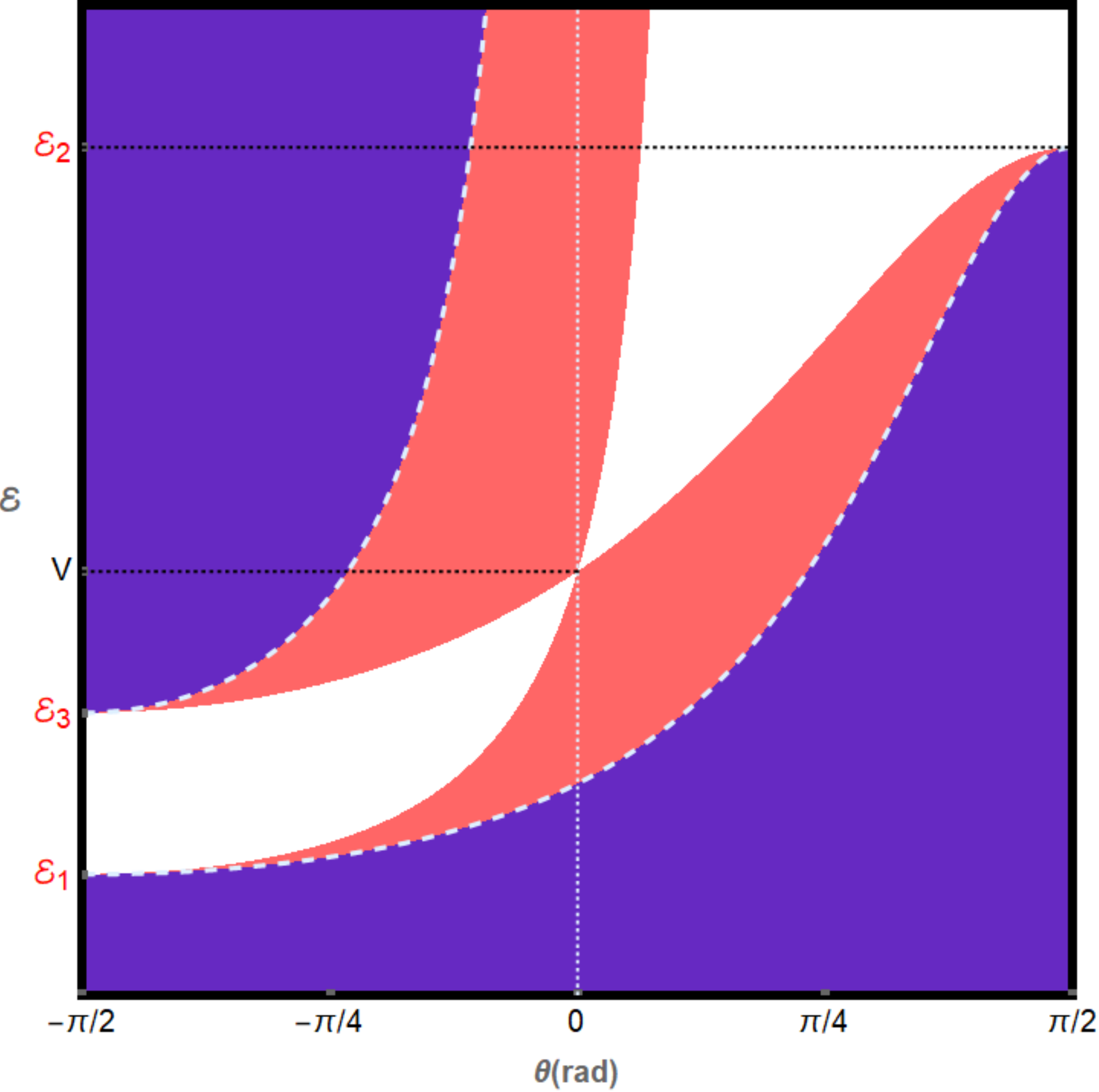}\label{TEnThetaPlusPi1:SubFigH}}
	\caption{(Color online) Comparative illustration of the refractive indices between the two regions of the system corresponding to the angles of incidence and refraction, depending on the energy $\epsilon$ and the tilt term $\tau $. The magenta zone represents when \( n_2 > n_1 \), and the pink zone represents when \( n_2 < n_1 \).}
	\label{TEnThetaPlusPi1}
\end{figure}

To assess and compare the refractive index between the tilted Dirac cone material and pristine graphene in our study, we plotted Fig. \ref{TEnThetaPlusPi1}, exploring all possible configurations by varying the values of \( \tau \), which is a key parameter in our analysis. This figure thus provides a precise visual representation of the refractive index between the two materials across different transmission ranges \cite{52,53}.
Using equation \eqref{001}, we plotted curves \( \epsilon = -\frac{V}{\tau  \sin \theta } \) and \( \epsilon = \frac{V}{2-\tau  \sin \theta } \) in the \( (\epsilon, \theta) \)-plane, along with their counterparts \( k_y = k_3 = -\frac{V}{\tau} \) and \( \epsilon = \frac{1}{2} (k_y \tau +V) \) in the \( (\epsilon, k_y) \) plane, where the refraction between the two media is equivalent \( ( n_2 = n_1 ) \). These curves are represented by dashed light blue lines. The magenta zones indicate \( n_2 > n_1 \), while the pink zones indicate \( n_1 > n_2 \), and the white zones are forbidden \cite{54,55,59}.
We observe that on the curves \( \epsilon=-\frac{V}{\tau \sin \theta } \) and \( k_y=k_3=-\frac{V}{\tau} \), we consistently witness the manifestation of Klein's paradox, where the transmission is total. To analyze and discuss the behavior of total transmission peaks at grazing incidence angles, we examined the variation of the transmission $T(\theta)$, as illustrated in Fig. \ref{TEnThetaPlusPi2}, for different values of $\tau$ near angles $-\pi/2$ and $\pi/2$ \cite{59}.

\begin{figure}[H]\centering
\hspace{-6mm}
	\subfloat[$\tau=0$]{\includegraphics[scale=0.182]{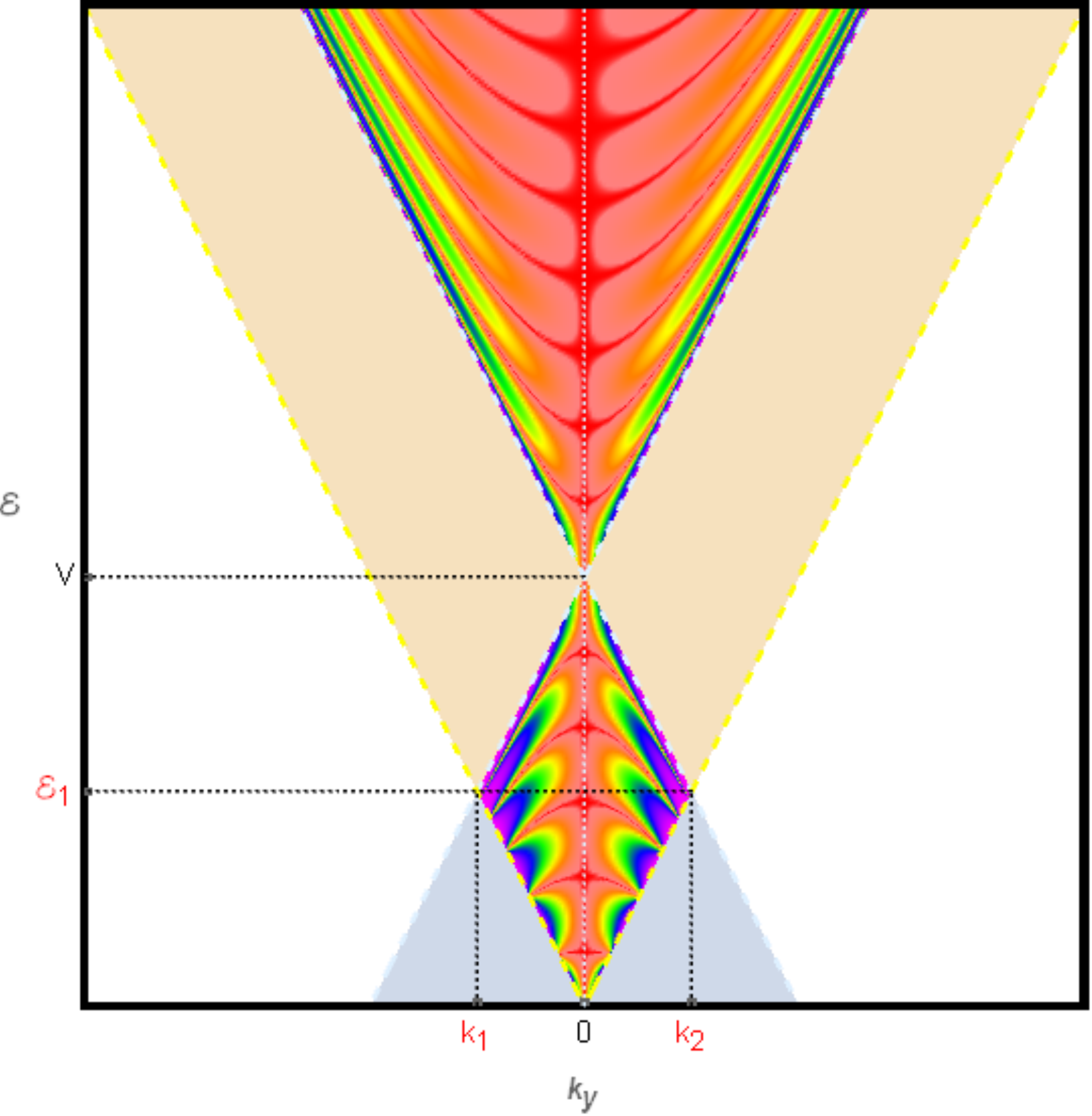}\label{TKytau:SubFigA}}
	\hspace{-1mm}
	\subfloat[$\tau=0.5 $]{\includegraphics[scale=0.182]{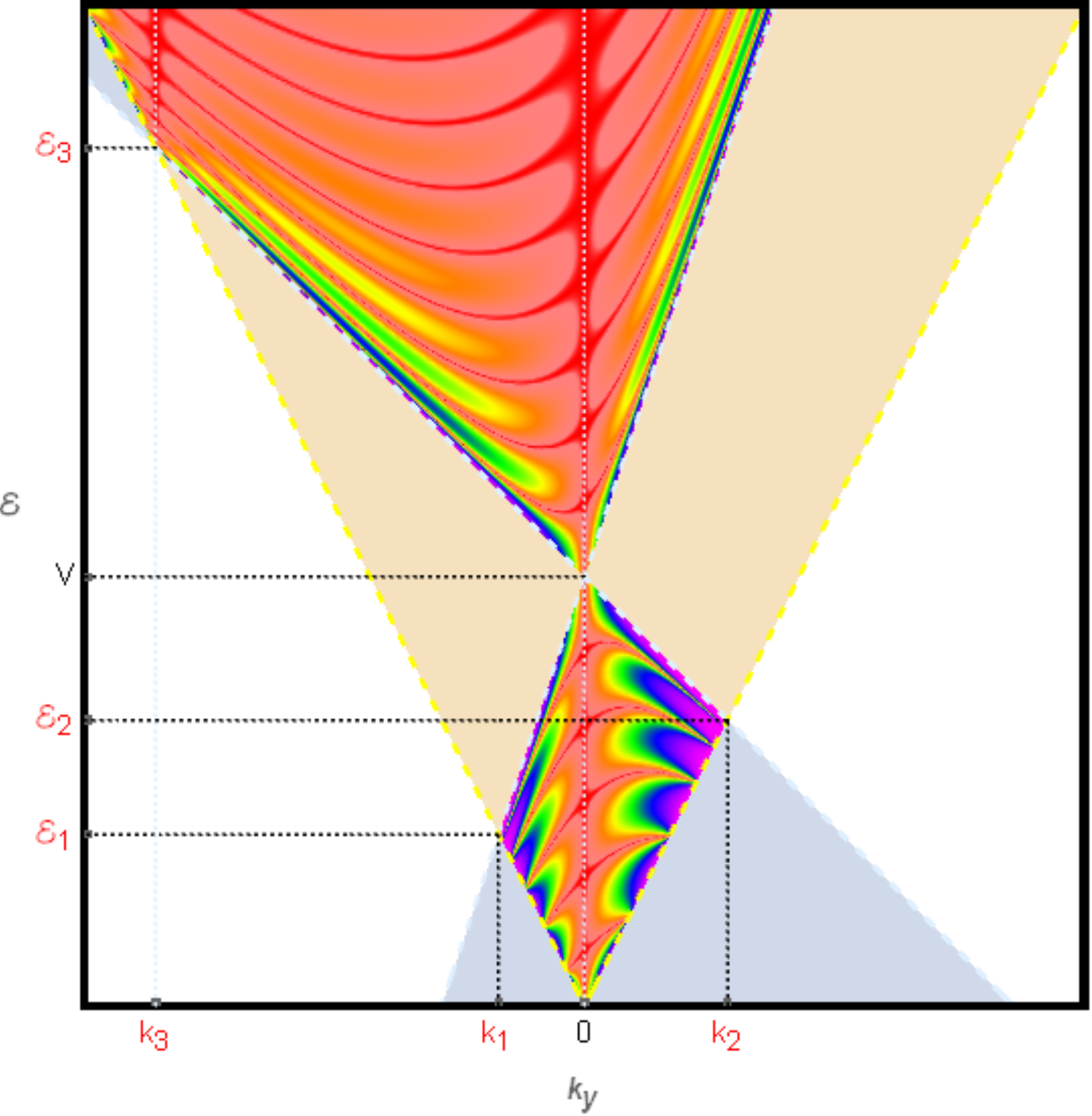}\label{TKytau:SubFigB}}
	\hspace{-1mm}	\subfloat[$\tau=1$]{\includegraphics[scale=0.182]{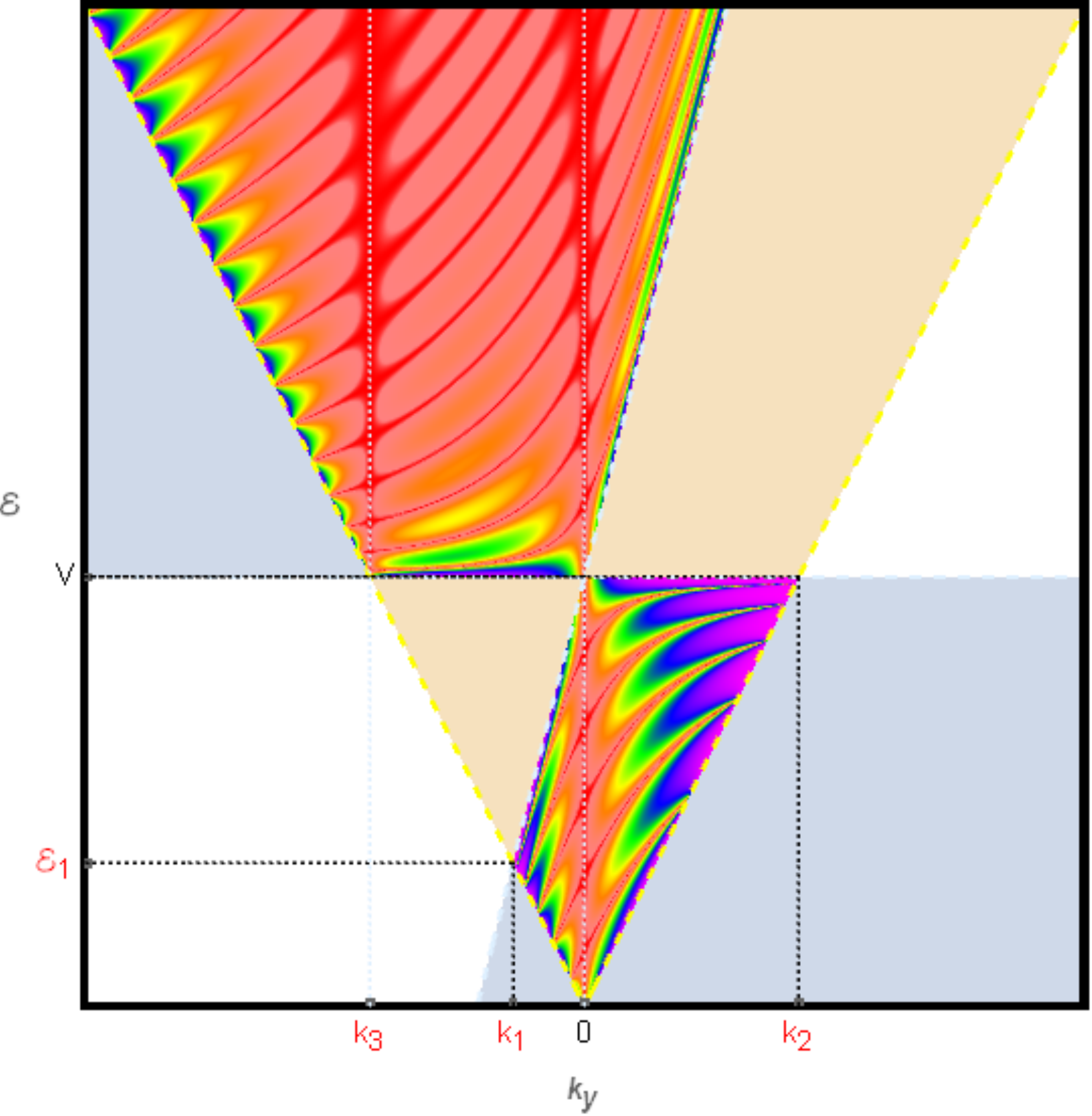}\label{TKytau:SubFigC}}
	\newline
	\subfloat[$ \tau=1.5$ ]{\includegraphics[scale=0.182]{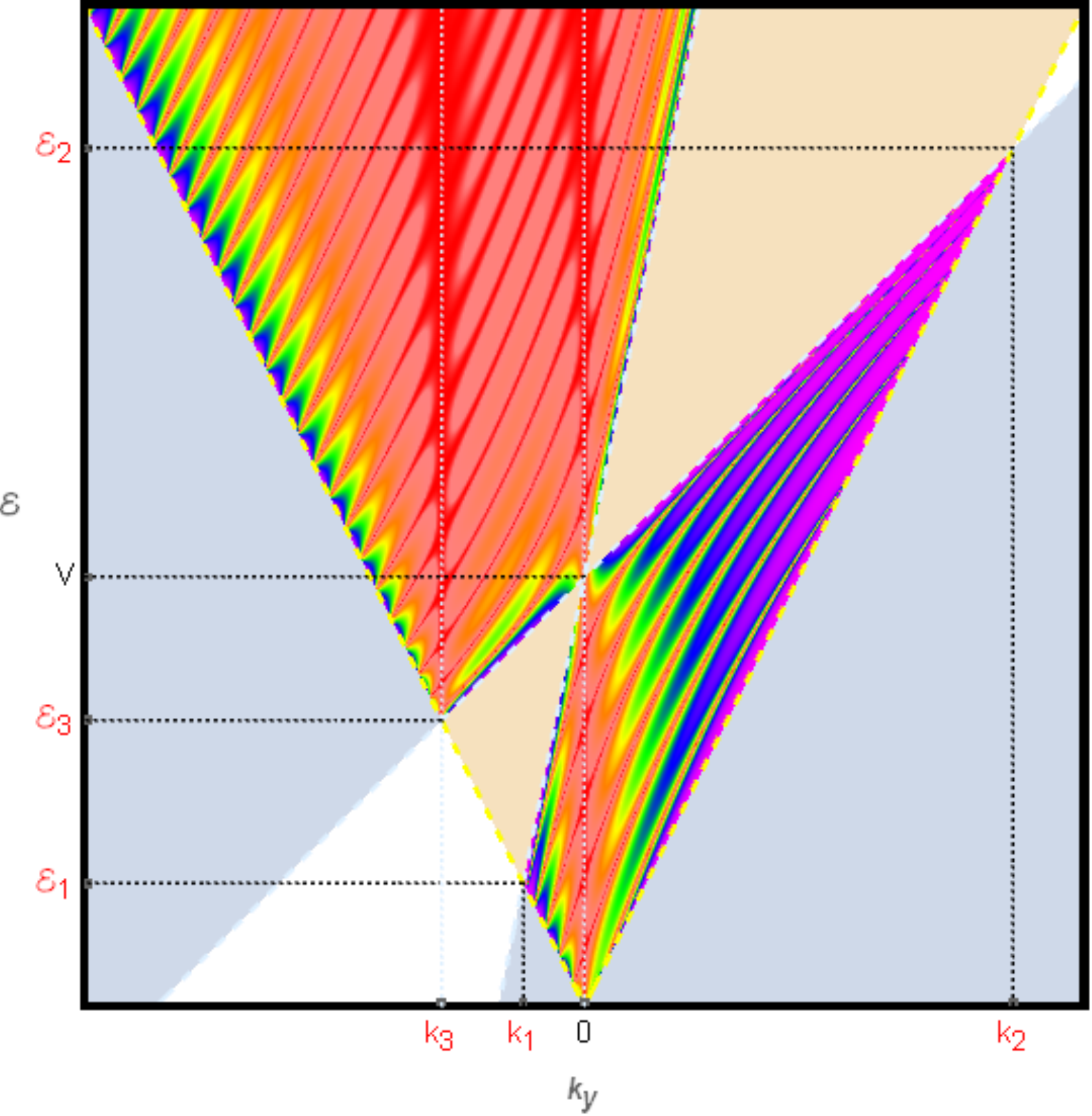}\label{TKytau:SubFigD}}
	\hspace{-1mm} \subfloat[$\tau=2$]{\includegraphics[scale=0.182]{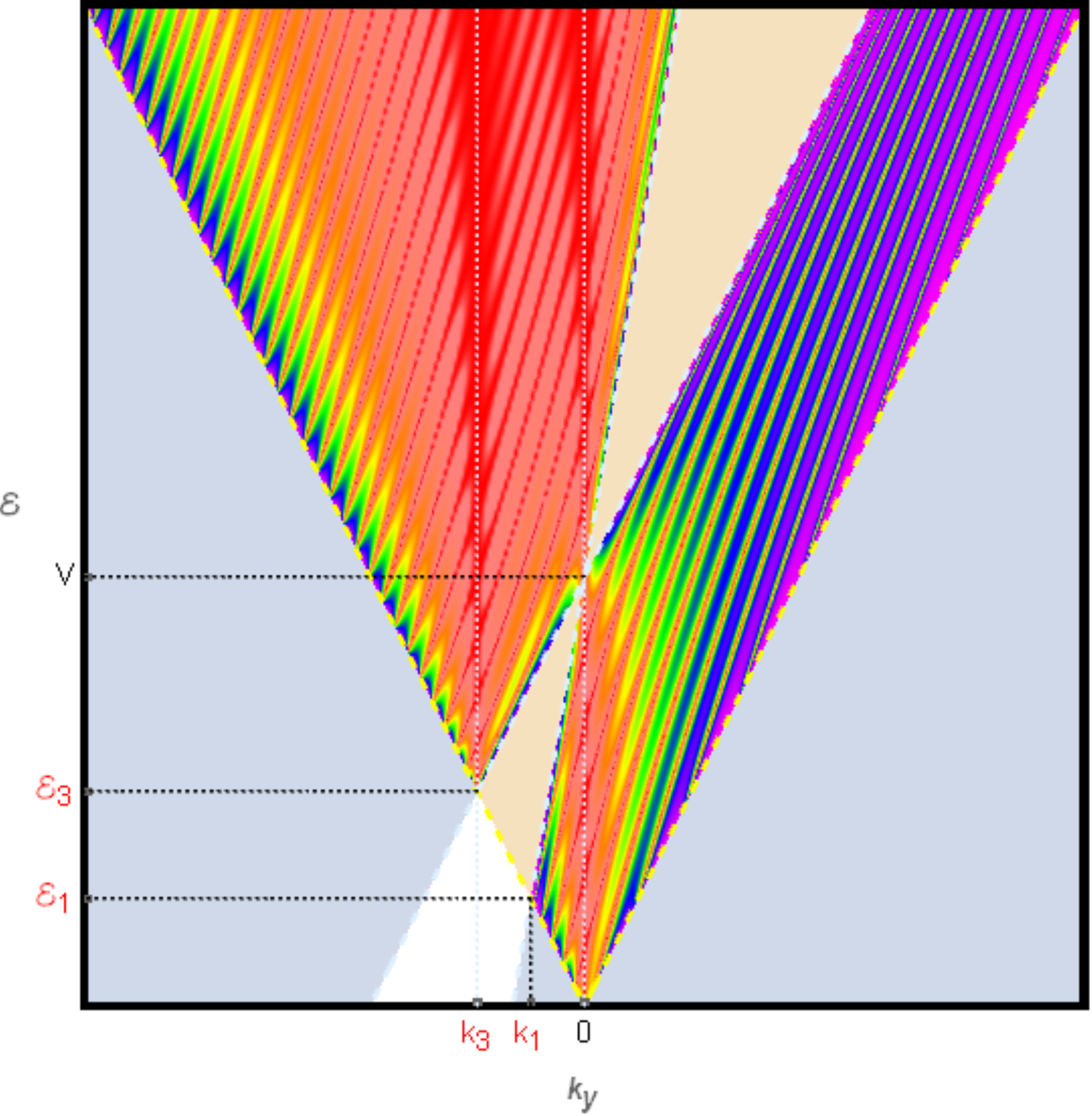}\label{TKytau:SubFigE}}
	\hspace{-1mm}
	\subfloat[$\tau=3 $]{\includegraphics[scale=0.27]{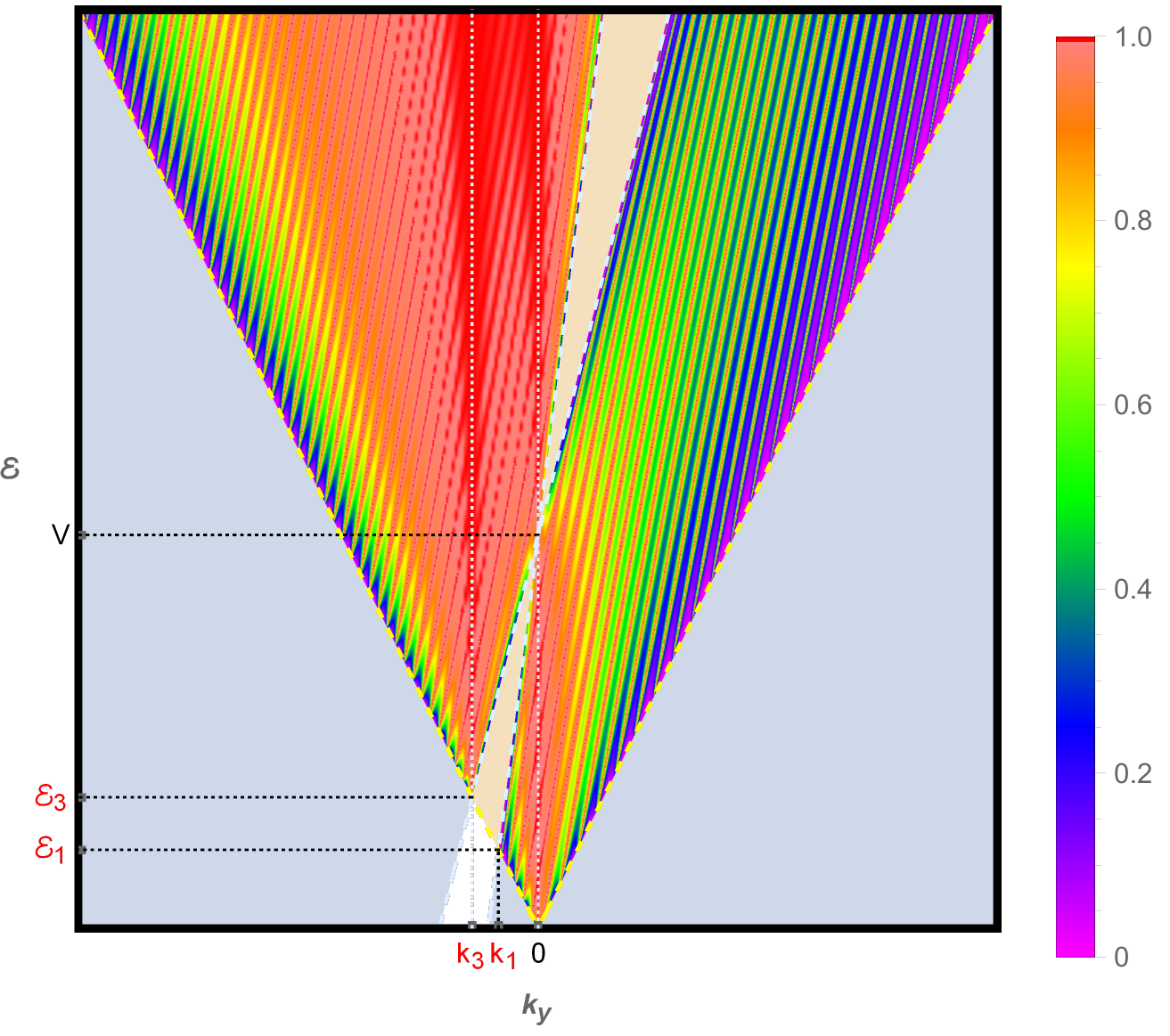}\label{TKytau:SubFigF}}
	\caption{(Color online) Density plot of transmission probability $T$ in ($\epsilon$, $k_y$)-plane for fixed value of barrier width $d=6\,\nano\meter$ and barrier height $V=3$ for six values of $\tau$. Incident energy $\epsilon$ is plotted in the $y$ axis and transverse momentum $k_y$ along $x$ axis.}
	\label{TKytau}
\end{figure}

In Fig. \ref{TKytau}, we have presented a plot depicting the transmission probability density $T$ in the plane defined by the energy $\epsilon$ and transverse momentum $k_y$. This plot was generated using specific values for the physical parameters involved. Specifically, we set the width of the barrier to be $d = 6$ nm, and the height of the barrier to $V = 3$. These parameters were chosen to represent typical values for the system under consideration. Additionally, we varied the tilt parameter $\tau$ across a range of values, specifically $\tau = 0, 0.5, 1, 1.5, 2,$ and $3$. By exploring this range of $\tau$, we aim to comprehensively understand its influence on the transmission probability density $T$. This detailed analysis allows us to gain insights into how different values of $\tau$ affect the transmission characteristics of the system and provides valuable information for further understanding and optimizing the performance of the heterostructure. The various cones, representing regions within the system, influence the transmission of Dirac fermions. The overlap of these cones gives rise to allowed and forbidden zones \cite{50,58,59}, as previously explained in terms of active surfaces. By expressing $k_x$ and $q_x$ as functions of $\epsilon$ and $k_y$ (Eqs \eqref{sa1} and \eqref{sa2}), we have obtained the different transmission density plots (Fig.  \ref{TKytau} ).

In Fig. \ref{TKytau:SubFigA} ($\tau=0$: Untilted configuration), the cones, depicted in yellow and blue, delineate forbidden zones represented by white areas and permitted zones consisting of a lozenge in the ($\epsilon,k_y$)-plane ($\epsilon < V$), where the tunneling effect is manifested, and a triangle corresponding to the scattering of Dirac fermions at the propagation energy ($\epsilon > V$). We notice that the transmission plot density exhibits an axial symmetry along the symmetry axis $ky=0$. On this axis, there is always a total transmition $T=1$ for any energy value, which corresponds to the Klein paradox, where the barrier becomes transparent to Dirac fermions. This paradox is also observed when total transmission peaks occur. These peaks take the form of parabolic branches that terminate at the two lower edges of the lozenge, whereas in the triangle, the branches tend towards infinity. The number of transmission peaks increases as a function of the barrier width $d$.
In the lower part of the lozenge, which is colored magenta ($\epsilon < V/2 $) (see Fig. \ref{TEnThetaPlusPi1:SubFigA}), the intermediate medium is more refractive than those of the two regions corresponding to the entry and exit of the system \cite{50,59}. Consequently, the angle of incidence $\theta$ is greater than the refracted angle $\phi$. Thus, for a grazing angle of incidence, there corresponds a critical angle in the intermediate region.

In Fig. \ref{TKytau:SubFigB} ($0<(\tau=0.5)<1$: type I), the overlapping of cones in this configuration shifts points $(k_1, \epsilon_1)$ and $(k_2, \epsilon_2)$ compared to the first configuration (Fig. \ref{TKytau:SubFigA}) and generates a third point $(k_3, \epsilon_3)$. This creates permitted transmission zones resulting from the overlap of yellow and blue cones (see Fig. \ref{TKytau:SubFigB}). Compared to the transmission density of the untilted cone, the transmission density loses its axial symmetry. In the deformed lozenge, the parabolic transmission peaks are conserved in number and are evenly distributed along the lower edges, resulting in a separation of these peaks on the right edge and a convergence on the left edge. The branches of peaks in the upper part of the density multiply and converge. In the right part of the zone, the parabolic branches of peaks tend towards infinity, while in the left part, they reach axis $k=k_3$ and then extend towards infinity. Along axis $k=k_3$, transmission is total, and the Klein paradox is also observed on this line, similar to axis $k=0$. The Fig. \ref{TEnThetaPlusPi1:SubFigB} highlights an important observation: when $\epsilon < (\tau k_y +V)/(2  +V)$ or $k_y <-V/\tau $ in the magenta zone, the intermediate medium is more refractive than the two regions corresponding to the entry and exit of the system. This difference in refraction results in the angle of incidence $\theta$ being greater than the refracted angle $\phi$. If $k_y >-V/\tau $ and  $\epsilon>(\tau k_y +V)/2 $, the region of the barrier is less refractive, and instead of discussing grazing angles, we talk about incidence critical angles \cite{49,59}.

The Fig. \ref{TKytau:SubFigC} depicts the transmission density for the case where ($\tau=1$: Type III). Upon examination, we observe that the density closely resembles that of the previous case. However, there are notable changes. Primarily, there is a discernible displacement of specific energy levels, denoted as $\epsilon_2=\epsilon_3=V$. This displacement leads to a distinct behavior characterized by an increase in the number of transmission peaks in the upper part of the density plot. Additionally, the alignment of specific energies results in the emergence of a line representing Dirac points.
Furthermore, it is worth noting that the peaks predominantly appear on the left side of the upper part of the plot. The shape of the lozenge transforms into a quadrilateral, with the upper right side being horizontal. The special quadrilateral of this case represents the limiting scenario where only the Klein paradox in transmission density is observed. It is interesting to observe that the lower right peaks move further apart, while the lower left peaks tighten. Overall, these detailed observations shed light on the nuanced changes in the transmission density and its characteristics in this particular configuration.
The Fig. \ref{TEnThetaPlusPi1:SubFigC}  designates the magenta zones as $n_2>n_1$ and the pink zones as $n_1>n_2$. In a similar manner to the previous scenario, if $\epsilon < \epsilon_1$, region $\textbf{\textcircled{2}}$ exhibits higher refraction compared to the input and output regions. For $\epsilon_1 < \epsilon < V$, as $k_y$ increases within the permitted zone, initially, regions $\textbf{\textcircled{1}}$ and $\textbf{\textcircled{3}}$ start by higher refractivity than region $\textbf{\textcircled{2}}$, but later, this refraction reverses at the expense of region $\textbf{\textcircled{2}}$. Furthermore, for $\epsilon > V$, when $k_y < k$, the second region becomes the most refractive, contrary to when $k_y > k$. However, when $k_y = k$, the system maintains the same refractive index, elucidating the Klein paradox along this axis.

Regarding ($\tau>1$: Type II) (Figs.  \ref{TKytau:SubFigD}-\ref{TKytau:SubFigF}), we have plotted three different cases: $\tau=1.5$, $\tau=2$, and $\tau=3$. This is because the coordinates of the point ($k_2,\epsilon_2$), defined as $k_2=\epsilon_2=V/(2-\tau)$, invert with the point ($k_3,\epsilon_3$). Specifically, the value $V/(2-\tau)$ tends toward infinity when $\tau=2$, while it becomes negative for $\tau>2$. These variations affect the behavior of the quadrilateral resulting from the lozenge found in the case of an untilted cone material (Fig. \ref{TKytau:SubFigA}).
When the parameter $\tau$ reaches 2 (Fig. \ref{TKytau:SubFigF}), a specific behavior occurs: the lower and upper right sides of the quadrilateral become parallel. This arises because the coordinates of point ($k_2, \epsilon_2$), defined by $k_2=\epsilon_2=V/(2-\tau)$, lead to a configuration where these sides of the quadrilateral align.
Furthermore, for values of $\tau$ greater than 2 (Fig. \ref{TKytau:SubFigF}), the lower and upper right sides intersect at a point with negative coordinates. This characteristic is also explained by the changes in the coordinates of point ($k_2,\epsilon_2$), where the value of $V/(2-\tau)$ becomes negative for $\tau>2$, thus resulting in this intersection at a point with negative coordinates.

As $\tau$ increases, the number of transmission peaks rises within the various permitted zones (Figs. \ref{TKytau:SubFigD}-\ref{TKytau:SubFigF}). Additionally, the Klein paradox persists along the $k_y=k_3$ axis due to the uniformity of refraction across the entire system (Fig. \ref{TEnThetaPlusPi1:SubFigD}).

\begin{figure}[H]\centering

	\subfloat[$\tau=0$]{\includegraphics[scale=0.15]{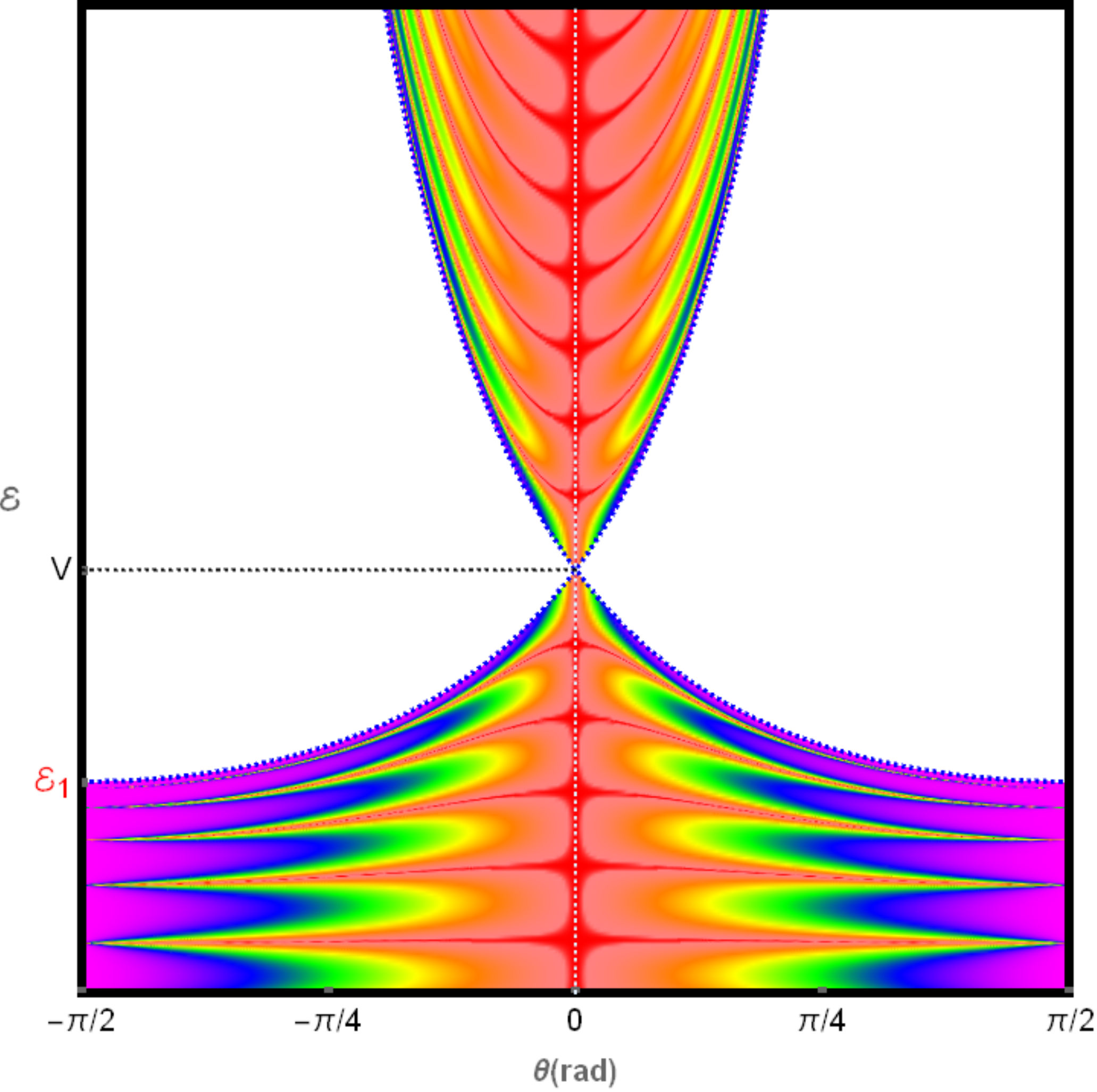}\label{Ttheta:SubFigA}}
	\hspace{-1mm}
	\subfloat[ $\tau=0.5 $]{\includegraphics[scale=0.123]{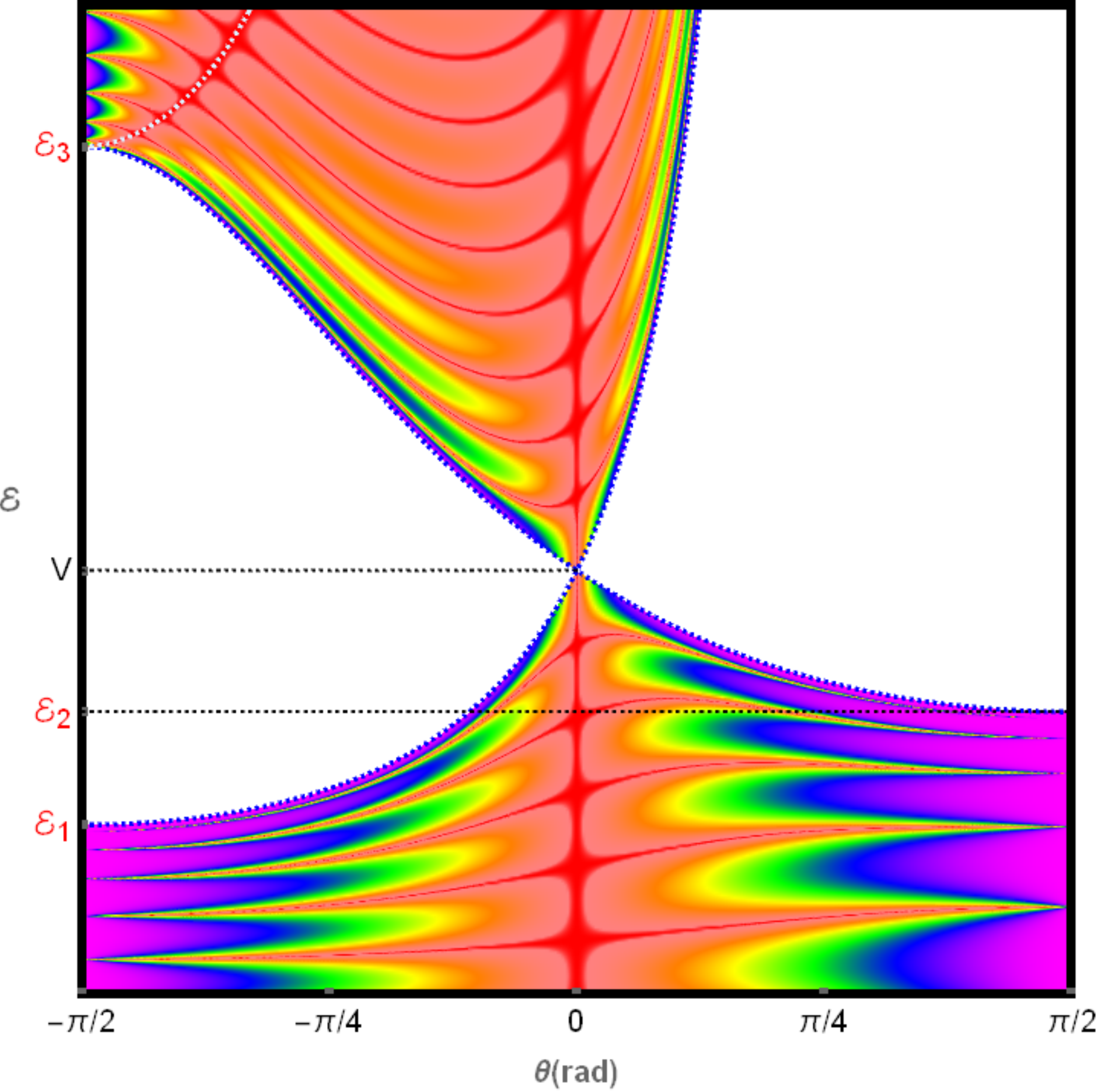}\label{Ttheta:SubFigB}}
	\hspace{-1mm}
	\subfloat[$\tau=1$]{\includegraphics[scale=0.123]{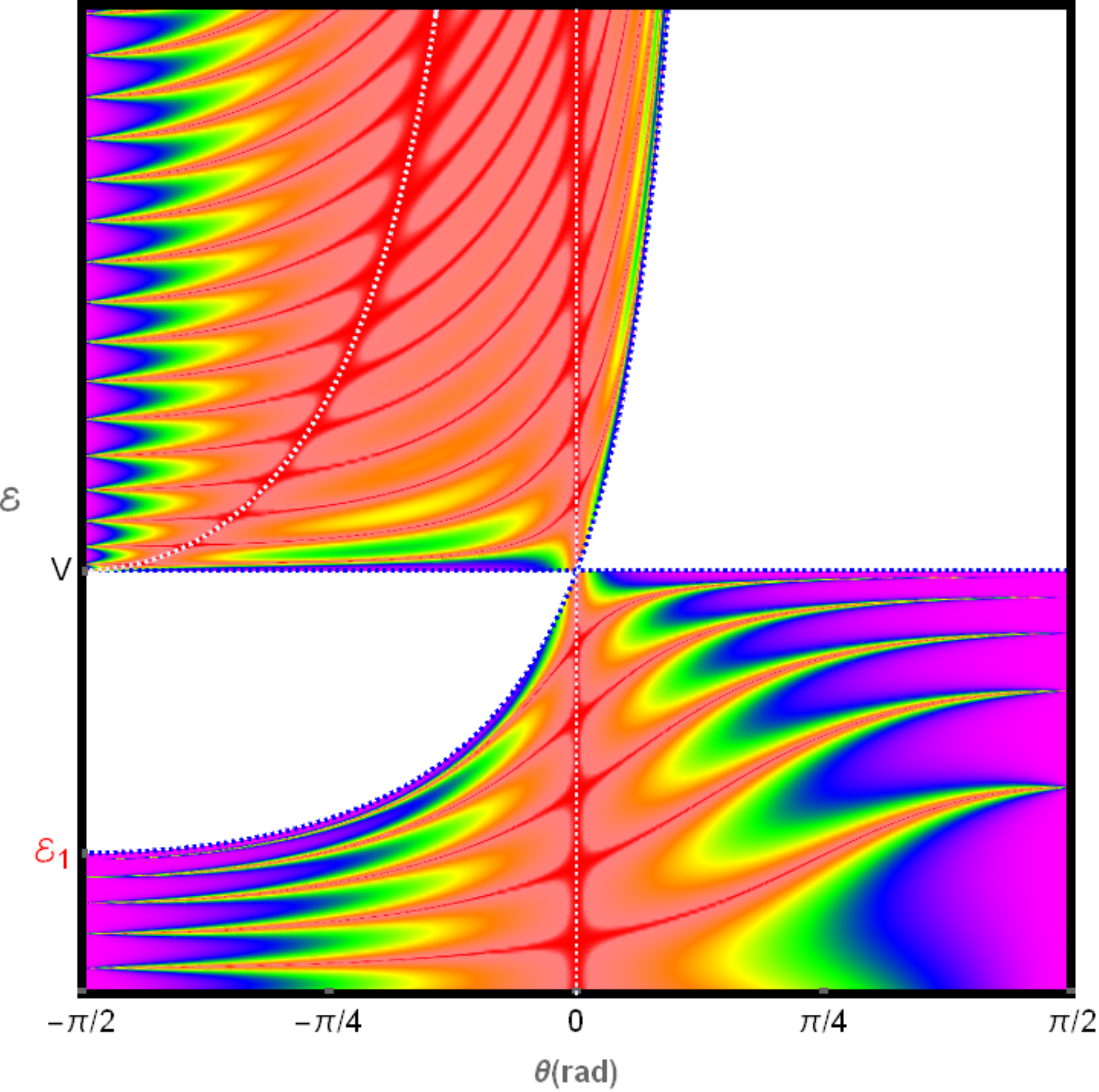}\label{Ttheta:SubFigC}}

	\subfloat[$ \tau=1.5$ ]{\includegraphics[scale=0.121]{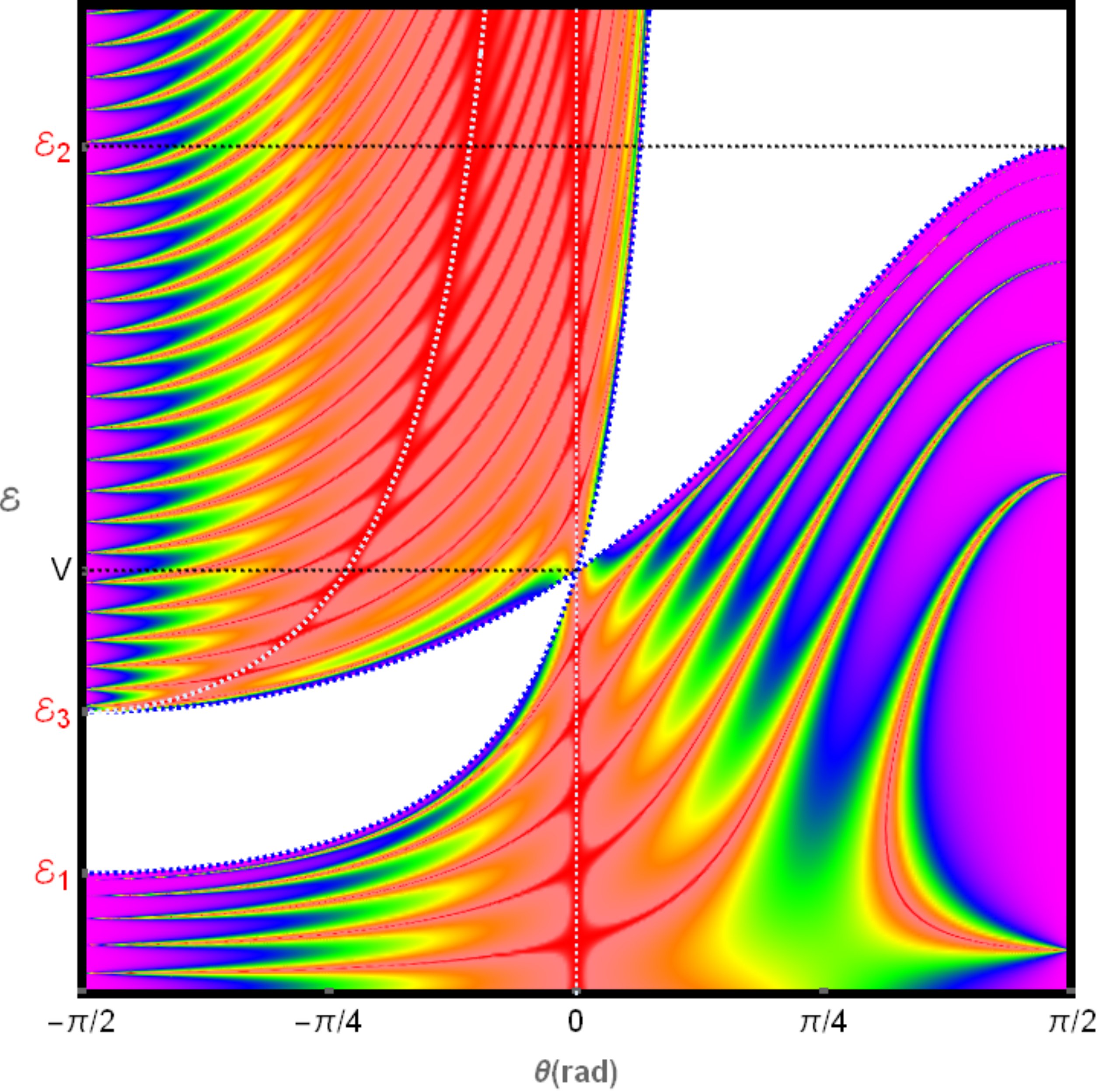}\label{Ttheta:SubFigD}}
	\hspace{-1mm}
	\subfloat[$\tau=2$]{\includegraphics[scale=0.121]{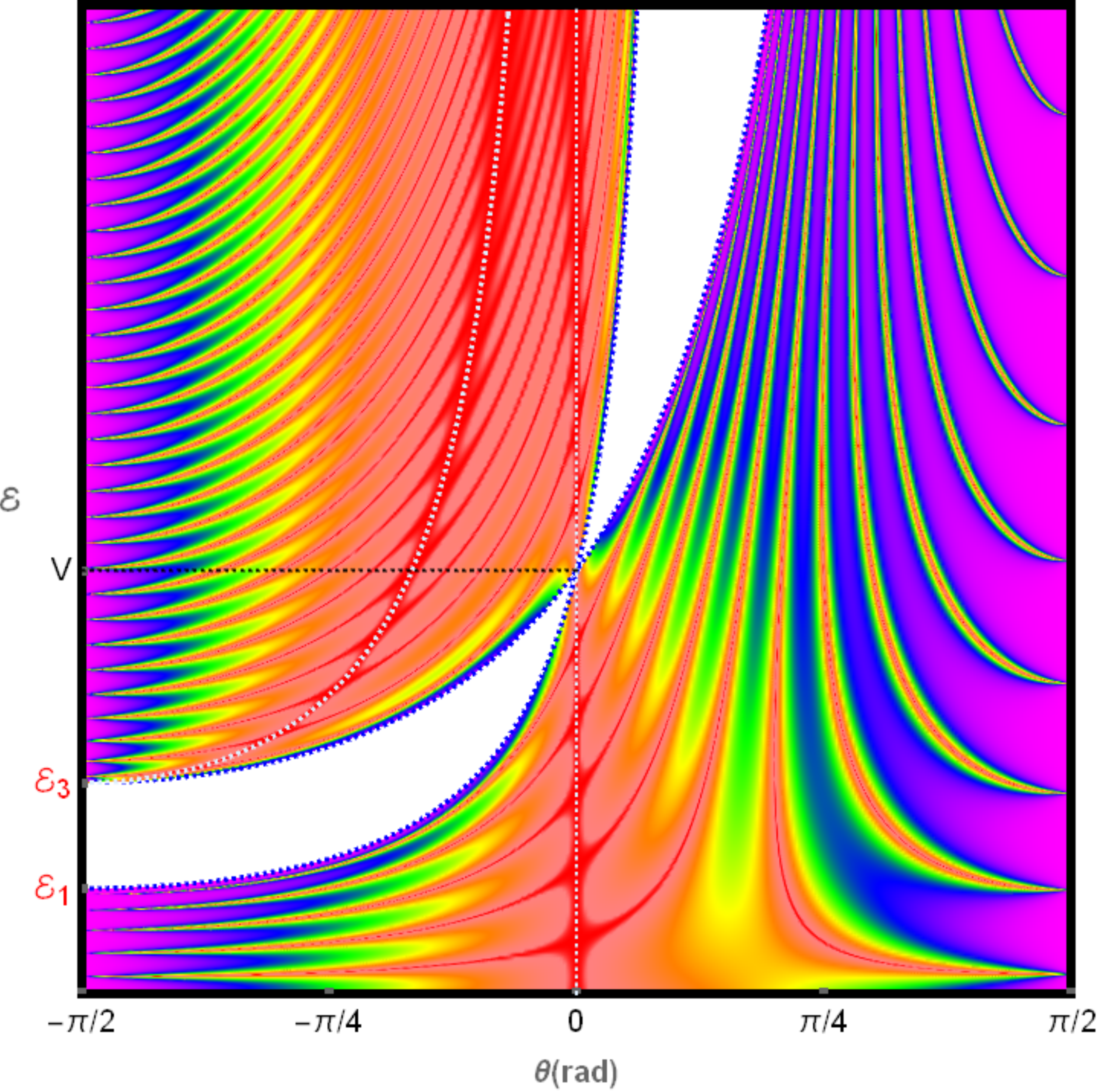}\label{Ttheta:SubFigE}}
	\hspace{-1mm}
	\subfloat[$\tau=3 $]{\includegraphics[scale=0.121]{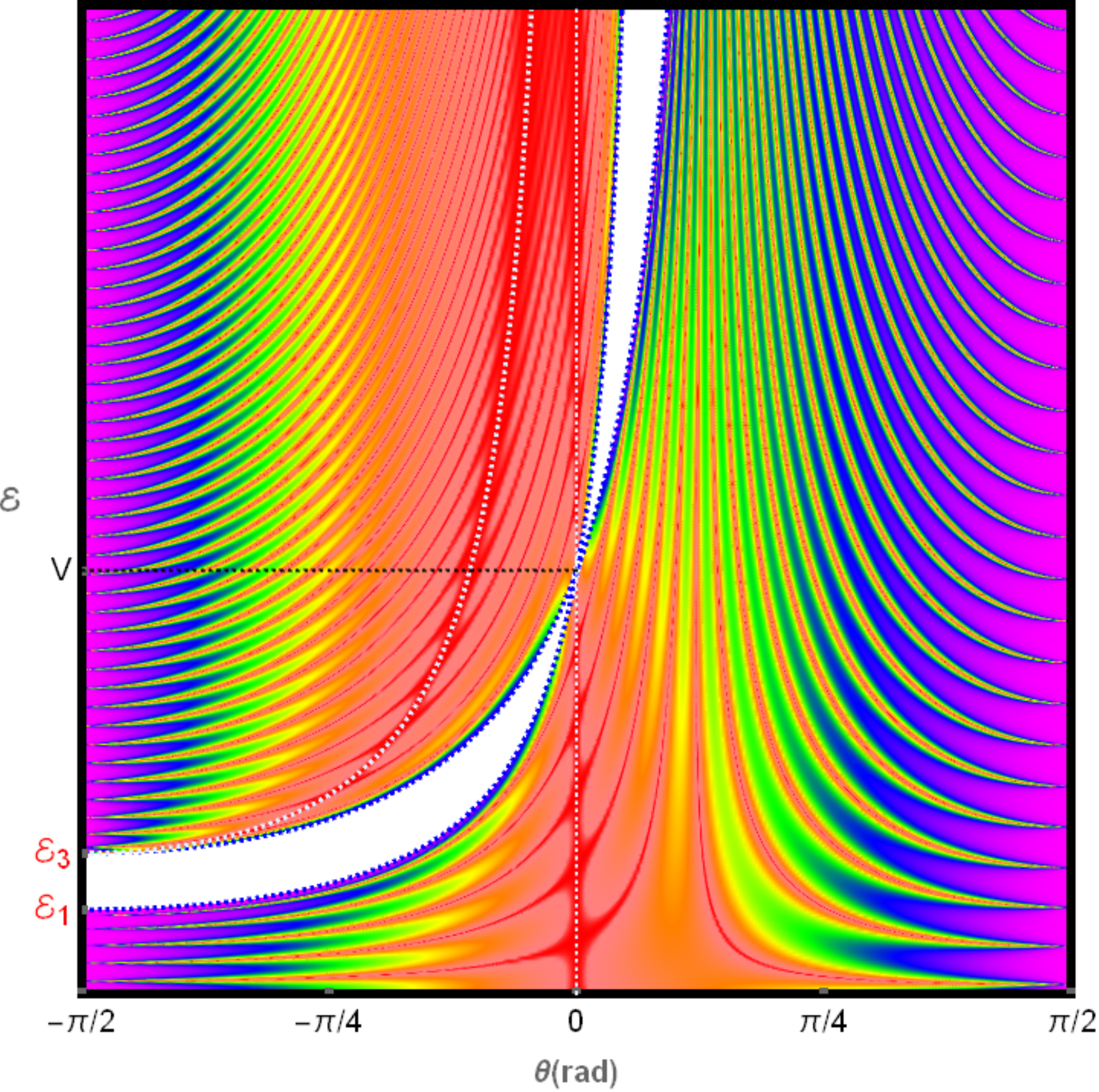}\label{Ttheta:SubFigF}}
	\caption{(Color online) Density plot of transmission probability $T$ in ($\epsilon$, $\theta$)-plane for fixed value of barrier width $d=6\,\nano\meter$ and barrier height $V=3$ for six values of $\tau$. Incident energy $\epsilon$ is plotted in the $y$ axis and incident angle along $x$ axis.}
	\label{Ttheta}
\end{figure}

Using the equation \eqref{sal1}, we can plot our transmission $T(\epsilon,\theta)$ results (Fig. \ref{Ttheta}) as a function of the incidence angle  at the junction between the first and second regions of the system.

This figure, similar to the one discussed earlier (Fig. \ref{TKytau}), presents density plots of the transmission $T(k_y, \theta)$ for various tilted Dirac cone configurations within the intermediate barrier region. The different permitted transmission zones in all these density transmission plots are delineated by the curves with equations $\epsilon=V/(1+(1-\tau)\sin\theta)$ and $\epsilon=V/(1-(1+\tau)\sin\theta)$, corresponding to cone interpenetration and thus to the previously discussed active surfaces. The white areas indicate forbidden transmission zones \cite{50,58,59}.

 Fig.  \ref{Ttheta:SubFigA} shows the density plot for ($\tau=0$: Untilted configuration), which exhibits axial symmetry at $k_y=0$ corresponding to normal incidence. Here, we observe the manifestation of the Klein paradox, which renders the barrier transparent with respect to the Dirac fermions diffusion.
For $\epsilon<V/2$, region $\textbf{\textcircled{2}}$ is the most refractive within the system (Fig. \ref{TEnThetaPlusPi1:SubFigE}). In this energy domain, all incident angles from $-\pi/2$ to $\pi/2$ are permitted, with each angle corresponding to a specific critical angle in region $\textbf{\textcircled{2}}$. This domain exhibits the same number of peaks as the  lozenge-shaped pattern in $T(\epsilon, k_y)$. These transmission peaks form a Dirac comb pattern, approaching the edges of the domain at grazing angles $\pi/2$ and $-\pi/2$, as clearly depicted in Figs. \ref{TEnThetaPlusPi2:SubFigA} and \ref{TEnThetaPlusPi2:SubFigE}.
From $\epsilon=V/2$, the refraction reverses (Fig. \ref{TEnThetaPlusPi1:SubFigE}), and the tunneling effect manifests in the domain between $\epsilon=0$ and $\epsilon=V$  similarly to what occurred in the lozenge on the $(k_y,\epsilon)$-plane. From $\epsilon=V$, we observe parabolic branches of transmission peaks that tend towards infinity.

In Fig. \ref{Ttheta:SubFigB} ($\tau=0.5$, representing Type I configurations), the active surfaces resulting from the collimation between the surfaces of different regions and those of the second region take the shape of an ellipse. This ellipse changes its focal length between positive and negative energies, generating a Dirac point at $\epsilon=V$. This Type I configuration results in a third point $(-\pi/2,\epsilon_3)$ at the boundaries of negative grazing angles, deforming the permitted transmission zones by producing additional peaks in this boundary for energies $\epsilon >\epsilon_3$ and tightening the peaks in the tunneling effect zone $\epsilon <V$, as shown in Fig. \ref{TEnThetaPlusPi2:SubFigB}. This shrinking of the peaks is due to the decrease of $\epsilon_1$. However, the transmission peaks on positive grazing angles ($\pi/2$) are moving apart, and $\epsilon_2$ is increasing. The number of peaks reaching the right and left grazing angles remains constant (see Figs. \ref{TEnThetaPlusPi2:SubFigB} and \ref{TEnThetaPlusPi2:SubFigF}).
 The Klein paradox effect is particularly observed in two distinct situations. Firstly, it occurs when the incidence is normal at the first junction of the system ($\theta=0$). Secondly, it also manifests when $\epsilon=-\frac{V}{\tau \sin \theta }$ for $\epsilon>V$, where the refractive index is constant, meaning $n_1=n_2$ in zones $\textbf{\textcircled{1}}$,$\textbf{\textcircled{2}}$  and $\textbf{\textcircled{3}}$, and the refringence remains uniform along the considered system. This latter condition is illustrated in Fig. \ref{TEnThetaPlusPi1:SubFigF}. It should be noted that for positive angles of incidence, the peaks of total transmission take the form of parabolic branches. These branches extend to infinity for energies $\epsilon>V$, and they become narrower compared to the untilted configuration. However, the peaks in the tunneling effect zone, in the region where $\epsilon<V$ moves away.
However, when the angles of incidence are negative, an interesting phenomenon occurs. The parabolic branches of the total transmission peaks extend from the normal incidence axis $\theta=0$ towards the curve defined by $\epsilon=-\frac{V}{\tau \sin \theta}$ for $\epsilon>V$. These branches then extend to the edge of the most grazing negative angles of incidence. However, in the region corresponding to the tunneling effect $\epsilon<V$, these branches narrow.

Fig. \ref{Ttheta:SubFigC} illustrates the transmission density plot of the Type III configuration ($\tau=1$), where the Fermi surfaces in region $\textbf{\textcircled{2}}$ of the system take the form of parabolas. At energy $\epsilon=V$, we observe the generation of a line consisting of a Dirac points set. The parabolas change their direction as they transition from negative energies to positive energies.
This Fig is similar to the previous one, except that the specific energies are equal to $\epsilon_2=\epsilon_3=V$, which results in the difference. This difference is reflected in the multiplication of branches in the total transmission peaks for energies $\epsilon>V$, the widening of branches for positive incidences, and their narrowing for negative incidences in the tunneling effect zone. All transmissions are allowed for energies $\epsilon<V$ in the case of positive incidences.
The peaks of total transmission for negative and positive grazing angles of incidence are illustrated respectively in Figs. \ref{TEnThetaPlusPi2:SubFigC} and \ref{TEnThetaPlusPi2:SubFigG}.

Figs \ref{Ttheta:SubFigD} to \ref{Ttheta:SubFigF} showcase the transmission density plots for $\tau=1.5$, $\tau=2$, and $\tau=3$, respectively. These plots offer insights into the diverse behaviors exhibited by Type II configurations. The differences observed in these plots stem from variations in the coordinates of the point $(\pi/2,\epsilon_2)$, which are determined by $(\pi/2,\epsilon_2)=(\pi/2, V/(2-\tau))$, as explained previously. In these three cases, we observe a very particular phenomenon: for the same energy level $\epsilon$ such that $\epsilon>\epsilon_3$, depending on the angle of incidence, the conduction of Dirac fermions changes nature from an electron in the conduction band to a hole in the valence band.
Fig. \ref{Ttheta:SubFigD} illustrates the conservation of branches of the total transmission peaks in the valence band for negative angles of incidence, while there is an increase in the number of peaks for positive angles of incidence.
\begin{figure}[H]\centering
\hspace{-2mm}
	\subfloat[$\tau=0$]{\includegraphics[scale=0.3]{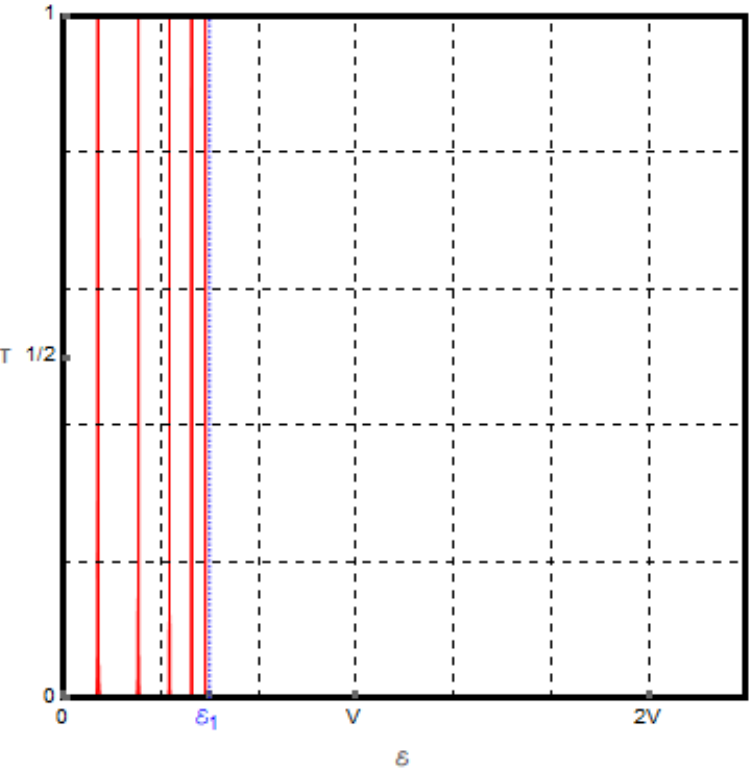} \label{TEnThetaPlusPi2:SubFigA}}
	\hspace{-0.5mm}
	\subfloat[ $\tau=0.5 $]{\includegraphics[scale=0.3]{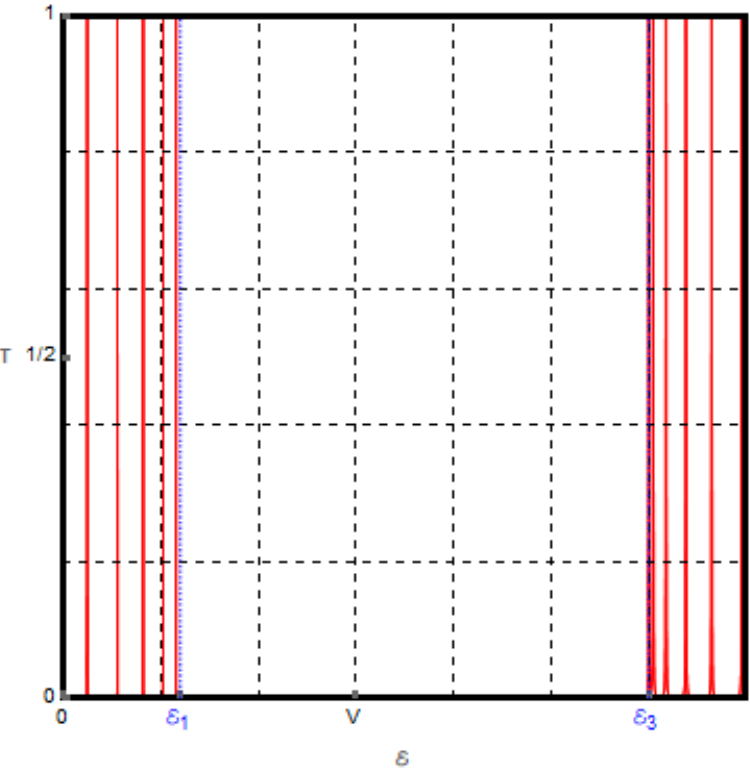} \label{TEnThetaPlusPi2:SubFigB}}
	\hspace{-1mm}
	\subfloat[$\tau=1$]{\includegraphics[scale=0.3]{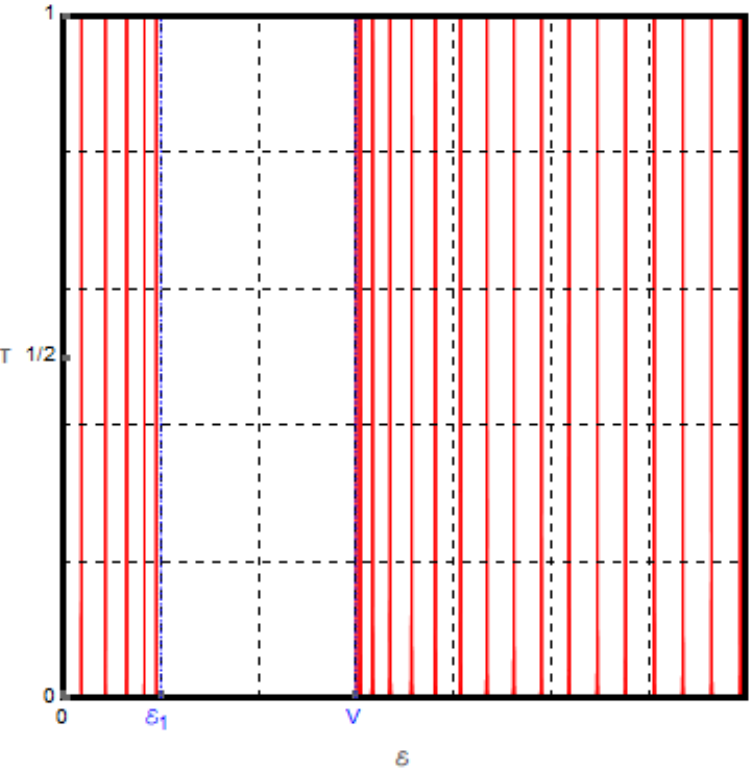} \label{TEnThetaPlusPi2:SubFigC}}
	\hspace{-0.5mm}
	\subfloat[$\tau=1.5$ ]{\includegraphics[scale=0.3]{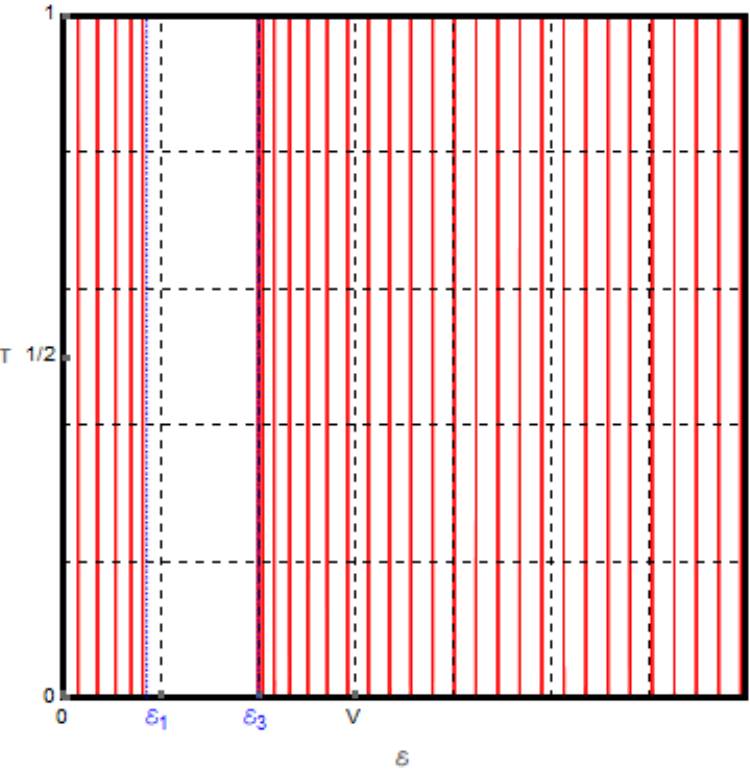}
\label{TEnThetaPlusPi2:SubFigD}}
\hspace{-3mm}
\subfloat[$\tau=0$]{\includegraphics[scale=0.3]{TEnThetaPlusPi2Tau00}
\label{TEnThetaPlusPi2:SubFigE}}
	\hspace{-1mm}
	\subfloat[ $\tau=0.5$]{\includegraphics[scale=0.3]{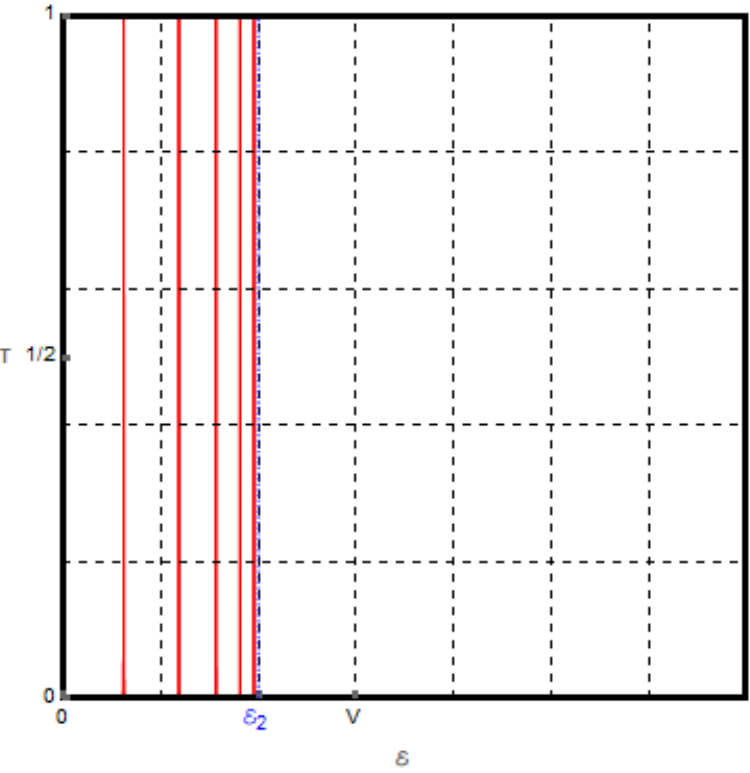}
\label{TEnThetaPlusPi2:SubFigF}}
	\hspace{-1mm}
	\subfloat[$\tau=1$]{\includegraphics[scale=0.3]{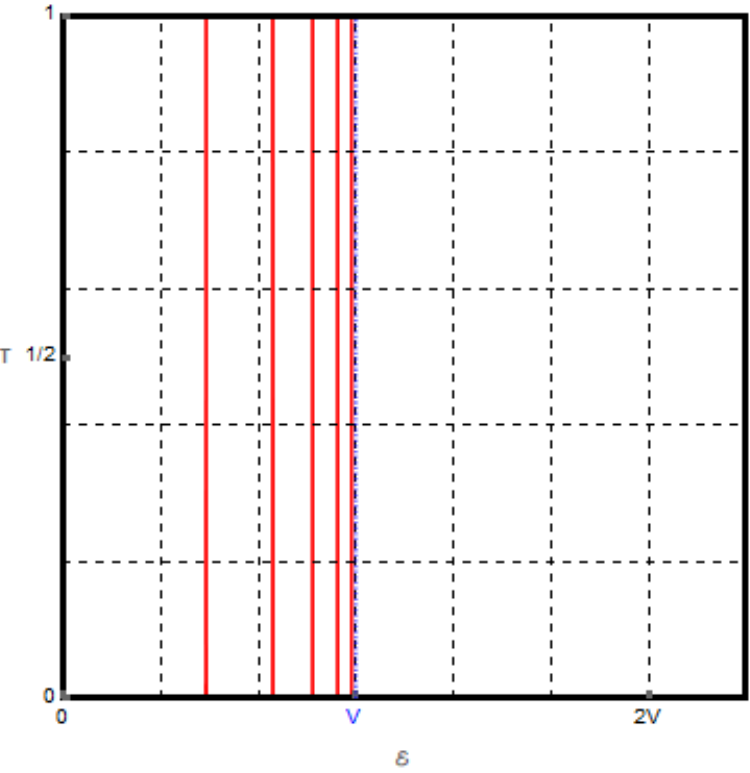}
\label{TEnThetaPlusPi2:SubFigG}}
	\hspace{-1mm}
	\subfloat[$ \tau=1.5$ ]{\includegraphics[scale=0.3]{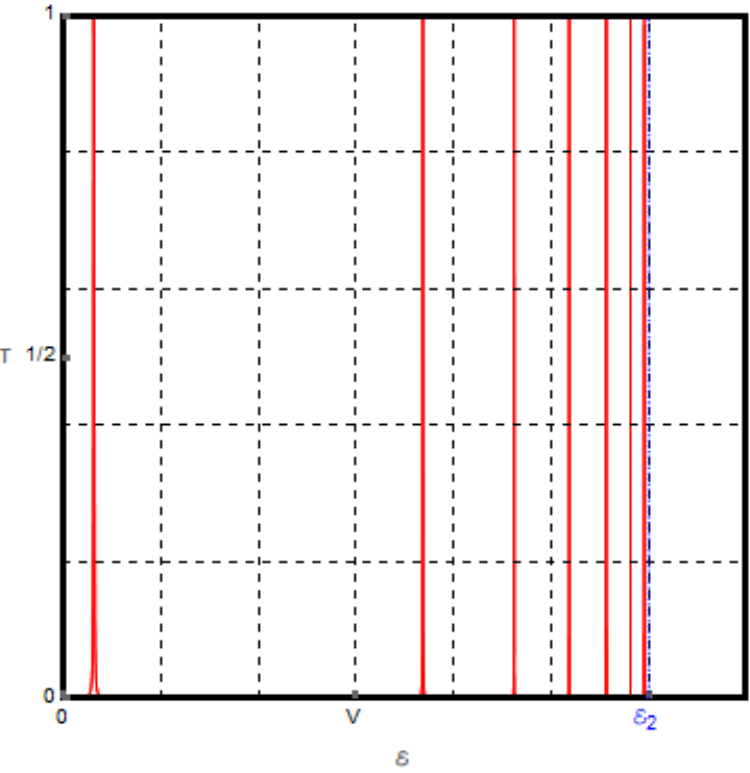}
\label{TEnThetaPlusPi2:SubFigH}}
	\caption{(Color online) The transmission probability $T(\epsilon)$ with respect to energy $\epsilon$ is plotted near the grazing incidence angles for a fixed barrier width of $d=4\,\nano\meter$ and barrier height $V=3$ for four values of $\tau=0, 0.5, 1,$ and $1.5$. Top: $\theta=-\tfrac{\pi}{2}$ and bottom: $\theta=\tfrac{\pi}{2}$.}
	\label{TEnThetaPlusPi2}
\end{figure}
Figs \ref{TEnThetaPlusPi2:SubFigD} and \ref{TEnThetaPlusPi2:SubFigH} respectively depict the peaks associated with negative and positive angles of incidence, illustrating the total transmission. These graphs allow us to observe the Dirac comb phenomenon of total transmission near grazing incidence angles. Additionally, Fig. \ref{TEnThetaPlusPi1:SubFigH} displays the refringence, providing supplementary information about the  properties of the studied system. In the conduction band, the peaks multiply and become closer together. As $\tau$ increases, the number of branches multiplies and converges in both the conduction and valence bands (see Figs \ref{Ttheta:SubFigE} and \ref{Ttheta:SubFigF}). From $\tau=2$, the branches of the total transmission peaks branch from the axis of normal incidence into the valence band on the side of the positive angles of incidence, extending up to l'infinity. On the other hand, the branches coming from infinity are flush with the transmission zone near the grazing positive angles of incidence.

To investigate the transmission behavior represented by $T(\epsilon, k_x)$, we delved into Eq. \eqref{eq4}, revealing the relation $k_y^2 = \epsilon^2 - k_x$. Given that the cone's tilt is along the $k_y$ axis, the system exhibits asymmetry concerning $k_y$. Consequently, a thorough analysis of our problem necessitates considering the direction of the transverse wave vector, specifically setting $k_y = m \sqrt{\epsilon^2 - k_x}$, with $m=\pm$. This ensures a comprehensive understanding of the system's behavior, accounting for its asymmetrical nature along the transverse wave vector direction.

\begin{figure}[H]\centering
	\subfloat[$0< \tau < 1, m=1$]{\includegraphics[scale=0.25]{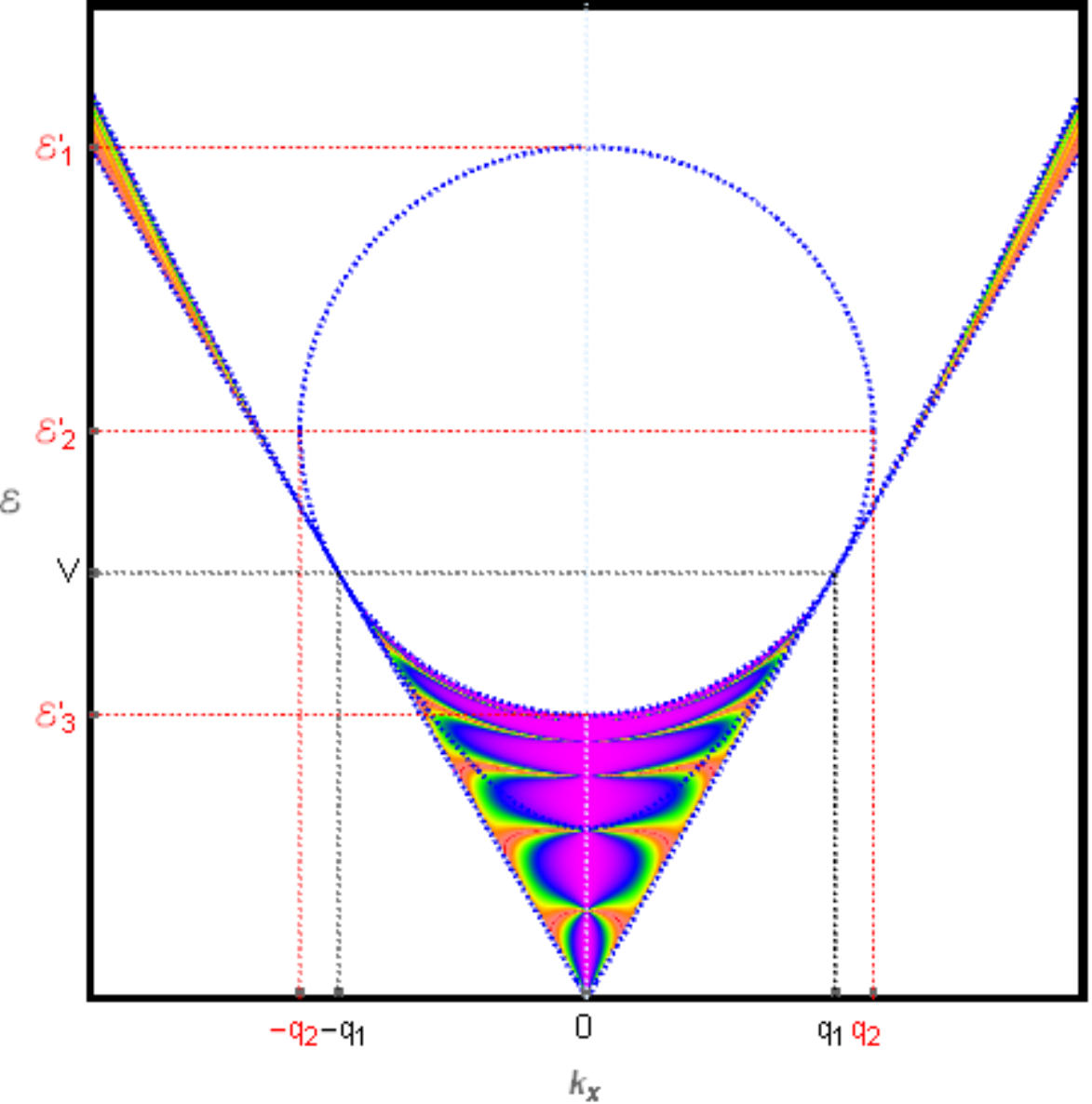}
\label{TKxtau:SubFigA}}
	\hspace{-1mm}
	\subfloat[$\tau=1, m=1$]{\includegraphics[scale=0.25]{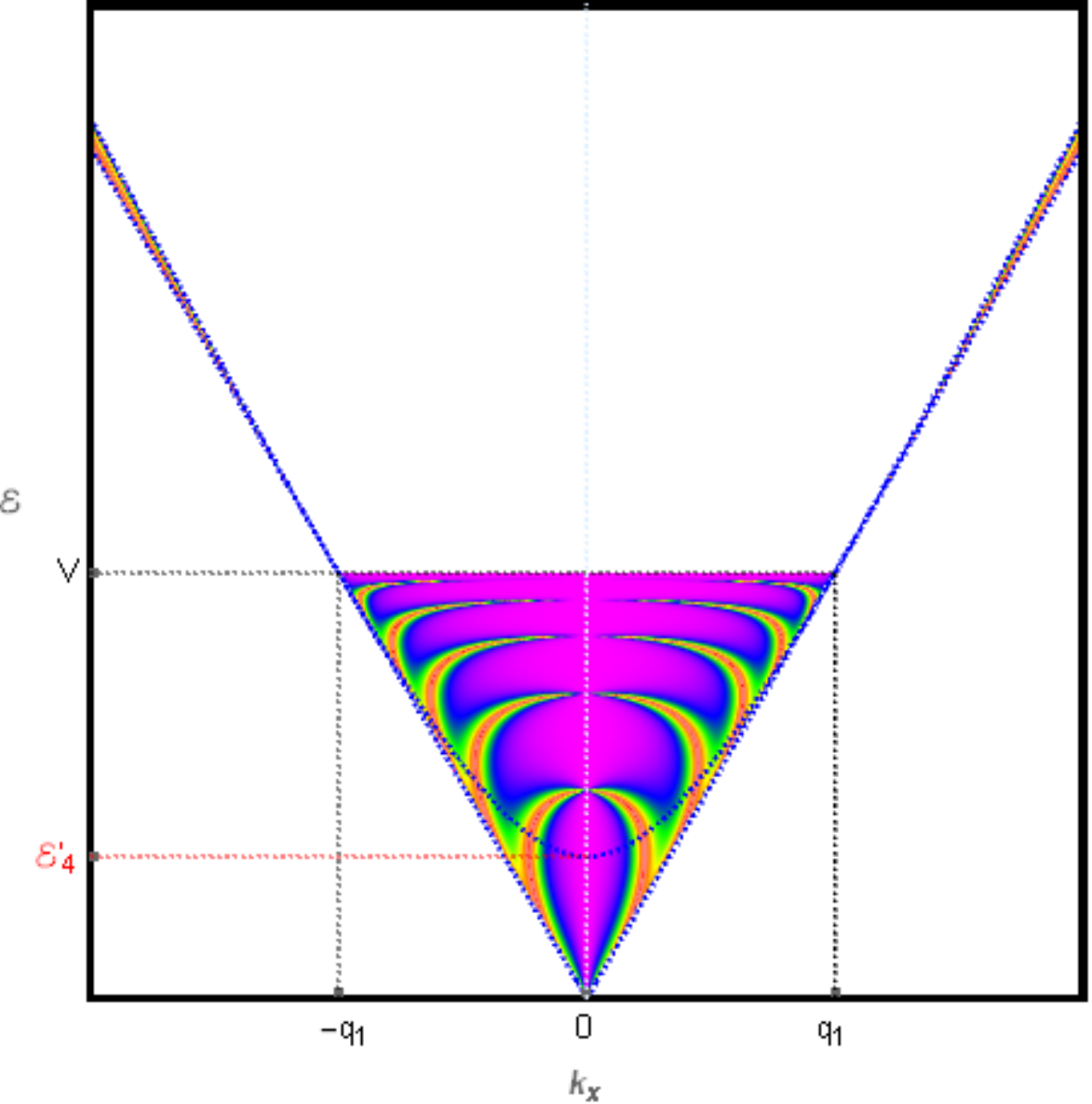}
\label{TKxtau:SubFigB}}
	\hspace{-1mm}
	\subfloat[$\tau=1.5, m=1$]{\includegraphics[scale=0.25]{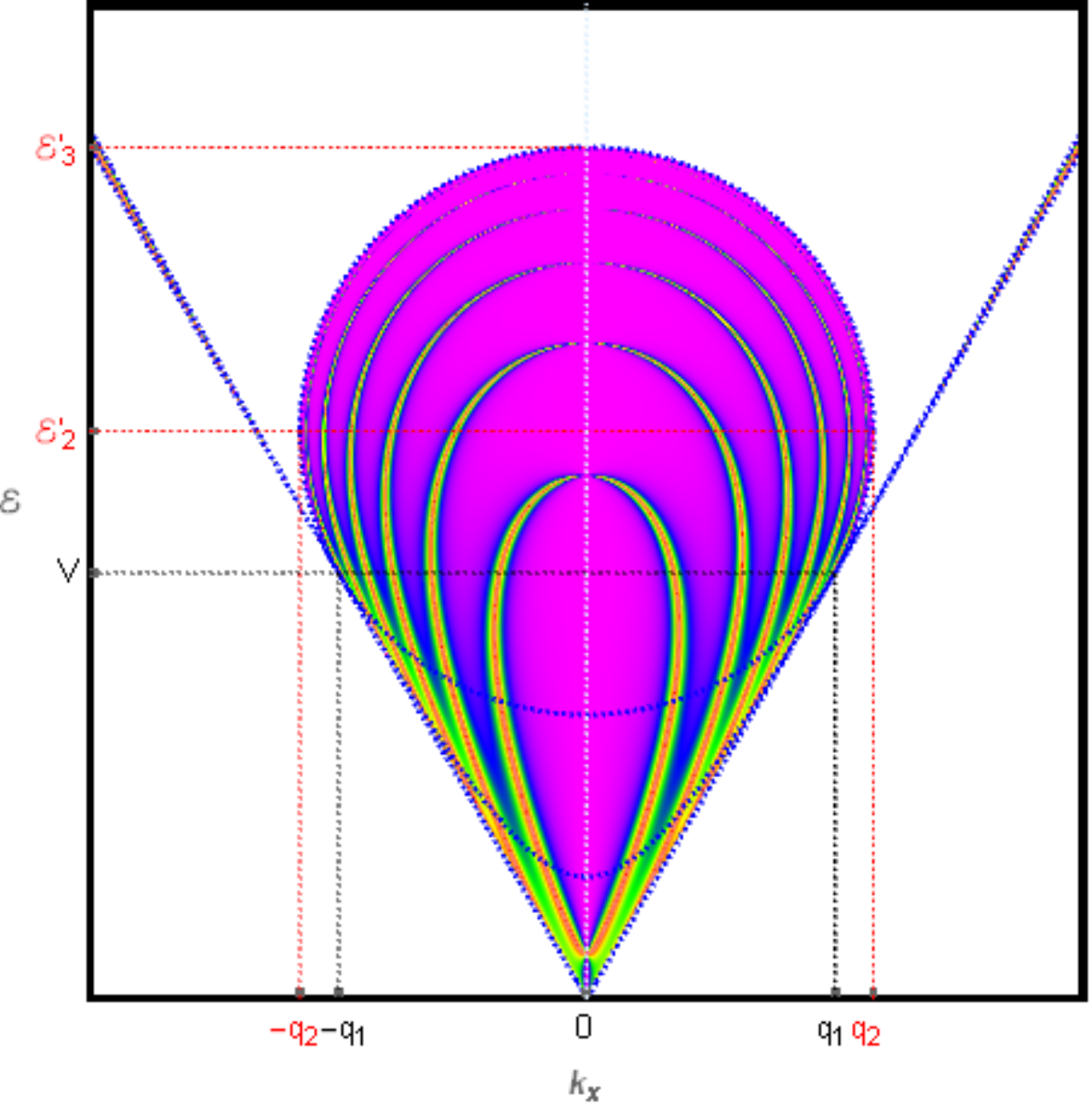}
\label{TKxtau:SubFigC}}
	\newline
	\subfloat[$ 0<\tau < 1, m=-1$ ]{\includegraphics[scale=0.25]{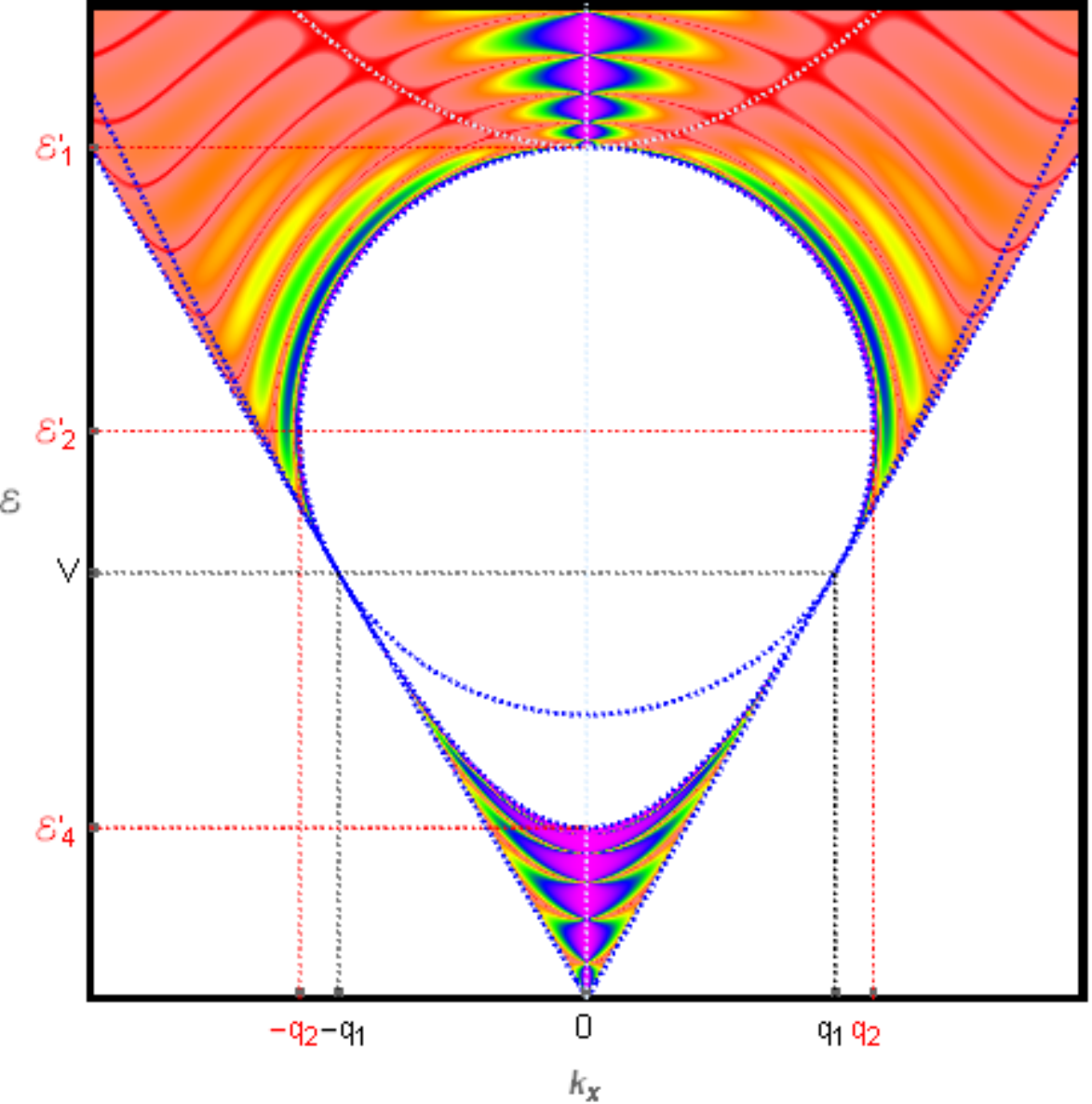}
\label{TKxtau:SubFigD}}
	\hspace{-1mm}	\subfloat[$\tau=1, m=-1$]{\includegraphics[scale=0.25]{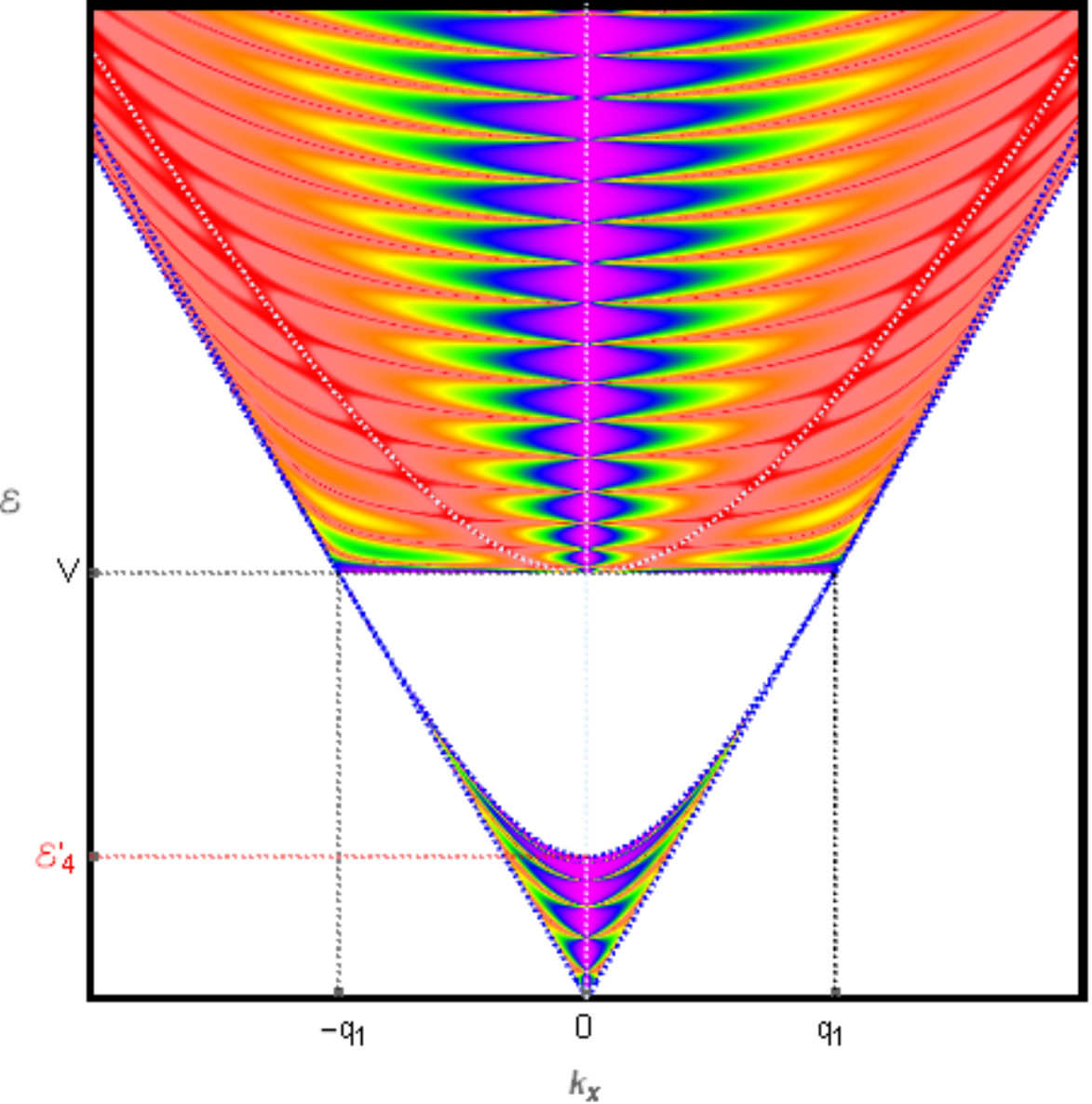}
\label{TKxtau:SubFigE}}
	\hspace{-1mm} \subfloat[$\tau=1.5, m=-1$]{\includegraphics[scale=0.25]{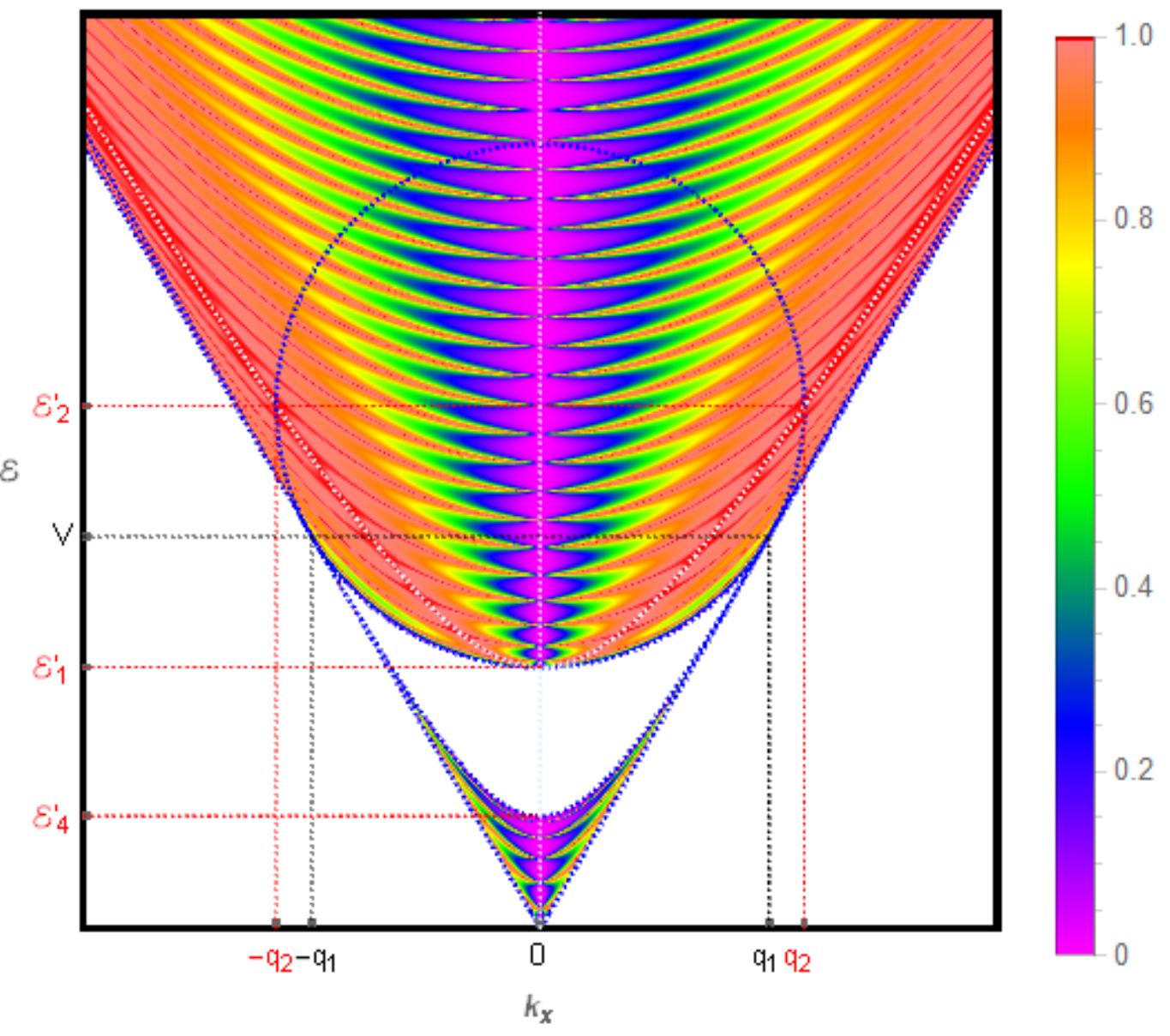}
\label{TKxtau:SubFigF}}
	\caption{(Color online) Density plot of the transmission probability $T$ in the ($\epsilon, k_x$)-plane for a fixed value of barrier width $d=4\mathrm{nm}$ and barrier height $V=3$ for three values of $\tau$. The incident energy $\epsilon$ is plotted on the $y$ axis, and the momentum $k_x$ is along the $x$ axis.}
	\label{TKxtau}
\end{figure}

Fig. \ref{TKxtau} presents a variety of density plots illustrating the transmission probability of Dirac fermions along the $x$ direction in the $(\epsilon, k_x)$-plane.  The figures at the top and bottom show the transmission density $T(\epsilon,k_x)$ for the positive direction $(m=1)$ and negative direction $(m=-1)$ of the transverse wave vector, respectively. A meticulous examination of the active surfaces in this plane has provided curves that delineate the different permitted and forbidden zones of the transmission density \cite{50,58,59}.

We have an ellipse, centered at the origin $(0, \epsilon'_2)$, is prominently featured, characterized by a major axis of length $2|q_2|$ and a minor axis of length $|\epsilon'_1-\epsilon'_3|$, with $\epsilon'_1=V/\tau$, $\epsilon'_2=V/(\tau (2 - \tau))$, $\epsilon'_3=V/(2+\tau )$, $q_1=V$, and $q_2=\frac{\sqrt{2} |\tau -1| V}{\sqrt{\tau (2 -\tau ) \left(2 \tau ^2-4 \tau +2\right)}}$. The equations governing the upper and lower half-ellipses are respectively given by:
\begin{align}
\epsilon &=-\frac{\sqrt{(\tau -1)^2 \left(k_x^2 (\tau -2) \tau +V^2\right)}+V}{\tau( \tau -2)  } \label{001}\\
\epsilon &=\frac{\sqrt{(\tau -1)^2 \left(k_x^2 (\tau -2) \tau +V^2\right)}-V}{\tau(\tau -2)} \label{002}
\end{align}
Additionally, a positive branch of hyperbola is described by:
\begin{equation}
\epsilon=\frac{\sqrt{(\tau +1)^2 \left(k_x^2 \tau (\tau +2)+V^2\right)}-V}{\tau (\tau +2 )} \label{003}
\end{equation}
with its vertex at $(0,\epsilon'_4)$, where  $\epsilon'_4=V/(2+\tau )$.
Lastly, a cone is defined by the two axes $\epsilon=k_x$ and $\epsilon=-k_x$.

First and foremost, it is essential to note that in each of the figures presented, the transmission of Dirac fermions is null as soon as $k_x$ equals zero. This observation is significant as it indicates that no propagation occurs when the wave vector component in the $x$ direction is zero. In the bottom figures (Fig. \ref{TKxtau}), on the parabolic branch defined by the equation $\epsilon=\sqrt{k_x^2+\frac{V^2}{\tau ^2}}$ (dashed light blue), the Klein paradox phenomenon is systematically observed, corresponding to a total transmission of Dirac fermions through the system. This behavior is explained by the fact that both regions of the system have the same refractive index. The peaks of total transmission branch out on either side of the parabolic branch. In all the figures, there is an axial symmetry with respect to the $k_x=0$ axis. Around this axis, total transmission peaks are observed, with their number increasing as $\tau$ increases. The areas where transmission is forbidden correspond to the white-colored zones \cite{50,58,59}.

For $\tau<1 $ $(\tau=0.5)$, it is observed that when $m=1$ and $\epsilon>V$, the permitted transmission zone lies between the cone and the positive hyperbolic branch, while for $\epsilon<V$, it lies between the cone and the lower half-ellipse (see Fig. \ref{TKxtau:SubFigA}). Conversely, for $m=-1$ and $\epsilon>V$, the permitted zone lies above the ellipse and within the cone, whereas for $\epsilon<V$, it lies between the cone and the positive hyperbolic branch (see Fig. \ref{TKxtau:SubFigD}).

In Figs \ref{TKxtau:SubFigB} and \ref{TKxtau:SubFigE}, where $\tau=1$, the ellipse flattens to form a segment at energy level $\epsilon=V$. For $\epsilon>V$ and $m=1$, the permitted transmission zone is similar to the case where $0<\tau<1$. However, for $\epsilon<V$, the permitted transmission zone is within the cone. On the other hand, for $\epsilon>V$ and $m=-1$, the permitted zone is also within the cone, but for $\epsilon<V$, it lies between the cone and the positive hyperbolic branch.

Figs \ref{TKxtau:SubFigC} and \ref{TKxtau:SubFigF} depict the transmission density $T(\epsilon, k_x)$ for $\tau > 1$ ($\tau = 1.5$), illustrating the cases where $m=1$ and $m=-1$, respectively. When $\epsilon>V$ and $m=1$, the permitted transmission region resembles that observed for $0<\tau<1$, with an additional area enclosed within the ellipse. However, when $\epsilon<V$, the permitted region lies entirely within the cone. Similarly, for $\epsilon>V$ and $m=-1$, the permitted region also lies entirely within the cone. If $\epsilon'_1<\epsilon<V$, the permitted region extends inside the ellipse. When $\epsilon<\epsilon'_1$, the permitted region lies between the cone and the positive hyperbolic branch.

%========================================================
\section{Conclusion}\label{Sec4}
%========================================================
After describing our system, composed of three regions including a barrier made of a simple tilted Dirac cone material, accompanied by an electrostatic potential that imposes a constraint on Dirac fermions circulating in graphene, we solved the Hamiltonians governing the Dirac fermions in each region of the system. We determined their spinors and eigenenergy spectra, connecting the various parameters of the regions while considering the conservation of the transverse wave vector. This allowed us to obtain a set of theoretical results rich in information about the system, including a relation similar to Snell-Descartes and Fermi surfaces that depend on the tilt parameter $\tau$.

In reciprocal space, we examined in detail the different configurations of conical geometric cones and Fermi surfaces for various energy levels. We also studied the Dirac points that appear at energy $V$, separating the conduction and valence bands, and then investigated the collimation of Fermi surfaces due to cone interpenetration to generate active surfaces where transmission is permitted. This transmission is due to the continuity of spinors at the interfaces between two consecutive regions of the system, explicitly stating the transfer matrix.

Following the theoretical study, we explored our analytical results in the form of figures to better understand the behavior of Dirac fermion transport through the system as a function of different physical parameters. We presented zones where the intermediate region is more or less refractive than graphene, as well as curves where we observe the Klein paradox, in addition to that of normal incidence. We also extensively discussed the transmission densities $T(\epsilon,k_y)$, $T(\epsilon,\theta)$, and $T(\epsilon,k_x)$ in terms of forbidden and allowed regions, as well as the effect of the Klein paradox, total transmission peaks, and refraction.
Our findings open avenues for the implementation of controllable electronic devices utilizing Dirac fermion collimation, governed by the tilt parameter, providing opportunities for precise manipulation and expanded functionality.

%========================================================

%=========================================================
\end{document}